# »The sun was darkened for seventeen days« (AD 797). An Interdisciplinary Exploration of Celestial Phenomena between Byzantium, Charlemagne, and a Volcanic Eruption

Johannes Preiser-Kapeller and Ewald Kislinger*

The blinding of the Byzantine emperor Constantine VI in Constantinople in August 797 and his overthrow by his mother Eirene, who then ruled as the first female »emperor« of the Eastern Roman Empire until 802, was used as legitimation for the coronation of the Frankish king Charlemagne as emperor of the Romans on 25 December 800, by contemporaries in Western Europe. Some observers in the West may have even interpreted the downfall of the Eastern Roman emperor and his replacement by a woman as sign of an impending collapse of the Roman Empire and the entire world order, as already expected (based on chiliastic cal- culations). We equally find indications of apocalyptic expectations in Constantinople, where Constantine´s blinding was linked with a spectacular celestial manifestation of divine dis- approval — a darkening of the sun for 17 days. In this paper, this obfuscation of the sun is compared with the description of other atmospheric and climatic phenomena in the 8th and 9th centuries, as well as before and after this period. In addition, natural scientific data is used to disprove earlier hypotheses on the physical background to this event and to present a more probable scenario (i.e., the impacts of one or more volcanic eruptions) for the dark- ening of 797 and other phenomena, which provided a peculiar »atmospheric« framework for the interpretation of the events between the downfall of Constantine VI and the coronation of Charlemagne by contemporaries.

* Correspondence details: Johannes Preiser-Kapeller, Institute for Medieval Research/Department for Byzantine Research, Austrian Academy of Sciences, Hollandstraße 11-13/4, 1020 Vienna, Austria, email: Johannes.Preiser- Kapeller@oeaw.ac.at. Ewald Kislinger, Department for Byzantine and Modern Greek Studies, University of Vienna, Postgasse 9/3, 1010 Vienna, email: Ewald.Kislinger@univie.ac.at.

*The Blinding of Emperor Constantine VI, the Reign of Empress Eirene and the Imperial Coronation of Charlemagne*

According to the *Chronographia*, a historical work written by the contemporary witness­es Georgios Synkellos and Theophanes,[1] on 15 August (or more probably 19 August) 797, Constantine VI,[2] son of Leon IV (r. 775-780)[3] and emperor of the Byzantine (Eastern Roman) Empire, was overthrown and blinded in Constantinople.[4] This mutilation was committed at least with the consent, if not by the order of his mother Eirene,[5] who became his successor. Her position was secured by Constantine´s maiming since the loss of his eyesight disqualified him for the throne. Eirene had already ruled for her underage son in the years 780 to 791, until conflicts emerged between mother and son when he reached the age of majority. Now she became the sole *basileus* (in Byzantine Greek: emperor).[6] Already the masculine form (*basileus*) indicates that Eirene, as the first female ruler of Byzantium, had a difficult stand­ ing in public opinion; additionally, she faced severe problems with regard to foreign affairs.[7] Her own overthrow in October 802[8] is to be seen as in some way connected with the cor­ onation of the Frankish king Charlemagne as emperor in Rome in 800[9], which called into question the sole claim of Byzantium (as »New Rome«) to the imperial title.[10] In turn, some Western sources were of the opinion that a woman could not be emperor and the imperial

---

Roman throne should therefore be considered vacant after the blinding of Constantine VI in August 797.[11] Eirene's position of power was already criticized in the Carolingian polemics against the Second Council of Nicaea in 787 in the *Libri Carolini*. There, Theodulf of Orléans not only quoted relevant passages from 1 Corinthians in which the silence of women in church affairs is decreed, but he even evoked the example of Ataliah, a queen of Israel in the Old Testament, who »justly« paid for her interference in the cult of the temple with her death. Looking ahead or looking back, after the blinding Constantine VI, an even more dramatic parallel to Eirene could have been identified, since Ataliah had ordered the murder of various members of her family.[12] Furthermore, in a letter to Charlemagne, the scholar Alcuin argued that after the blinding of the emperor in Constantinople and of Pope Leo III in Rome (in 799), only he (Charles) was left to preserve the »threatened salvation of the churches of Christ« (*tota salus ecclesiarum Christi*).[13]

Also drawing inferences from Alcuin´s statement, scholars such as Richard Landes have hypothesized that the information about Constantine´s fall arrived at Charlemagne´s court in an atmosphere already full of apocalyptic fears. Based on older calculations that the creation of the world could be dated to 5199/5200 BC, some contemporaries in the Latin Christian West expected the dawn of the 7th millennium of the world in the year AD 800 and thus the end of times. Landes assumes that in the Frankish Empire, before and during the reign of Charlemagne (768-814), the meaning of the year 6000/800 was deliberately concealed in textual tradition. Tacitly, however, the conversion to a chronology after the birth of Christ was promoted to avoid the potentially disastrous turn of the millennium. Accordingly, the date 25 December 800 was selected for Charles's coronation as emperor, symbolizing the assurance of the continued existence of the Roman Empire and world order.[14] Against this background, the »coup d'état« of Empress Eirene in 797 would have been interpreted by Charlemagne and his entourage as another sign of the impeding fall of the Roman Empire (construed according to 2 Thessalonians 2.6 as the power that holds back the dawn of the last days). Out of this grew a necessity to renew the Roman Empire in the West, and hence Charlemagne´s coronation.[15] Landes´ interpretation is not universally accepted in scholarship. A controversy exists concerning how widespread such apocalyptic assumptions were among the scholars of Latin Europe or in the immediate retinue of Charlemagne.[16] Nevertheless, it is intriguing that the blinding of Constantine VI was likewise related to a symbolic manifestation of divine volition in Constantinople.[17] The *Chronographia* reports for the Byzantine year AM (annus mundi) 6289 (1 September 796-31 August 797), that after Constantine's blinding the sun did not shine in its usual ways for 17 (ιζ´) days.[18] This the people of Constantinople considered a sign of celestial disapproval. The *Chronographia* says:

> About the 9th hour they blinded him [Constantine VI] in a cruel and grievous manner with a view to making him die at the behest of his mother and her advisers. The sun was darkened for seventeen days and did not emit its rays so that ships lost course and drifted about. All said and agreed (καὶ πάντας λέγειν καὶ ὁμολογεῖν) that the sun withheld its rays because the emperor had been blinded. In this manner his mother Eirene acceded to power.[19]

---

11 *Annales Laureshamenses*, ed. Pertz, 38: *tunc cessabat a parte Graecorum nomen imperatoris, et femineum imperium apud se habebant* (»and because at that time the imperial dignity had ceased on the part of the Greeks and they had female rule«).

12 *Opus Caroli regis contra synodum* III, 13, ed. Freeman and Meyvaert, 385-387. See also Noble, *Images, Iconoclasm, and the Carolingians*, 158-206, esp. 185-190.

13 *Epistolae Karolini Aevi* 2, no. 174, ed. Dümmler, 288. See also Höfert, *Kaisertum und Kalifat*, 396.

14 Landes, Lest the Millennium Be Fulfilled, esp. 181-203.

15 Landes, Lest the Millennium Be Fulfilled, 201: »Just as Augustine and Jerome's contemporaries could see Rome's sack through the lens of II Thessalonians, so too could Alcuin and Charlemagne view a woman's usurpation of the imperial throne in Constantinople as the signal that the Fourth Empire had fallen and that Antichrist had been unleashed—and, in fact, Eirene's coup occurred in 797 AD, or, according to AM II, in the last five years of the final millennium.«

16 See also Fried, Endzeiterwartung, esp. 396-398; Brandes, »Tempora periculosa sunt«, 49-79; Möhring, *Der Welt- kaiser der Endzeit*, 25-26, 136-143; Schieffer, *Neues von der Kaiserkrönung*, 24; Palmer, *Apocalypse in the Early Middle Ages*, 4-21 (summing up the debate) and 130-158 (discussing the theories of Landes and Brandes and of their critics); Fried, *Karl der Große*, 437-439, 462-465, and Wozniak, *Naturereignisse im frühen Mittelalter*, 713-716.

Some Byzantine apocalyptic texts, which were probably composed during Eirene´s reign, may suggest an even more intense debate and negative interpretation of the emperor´s blinding and Eirene´s ascent to power. These texts took up earlier prophecies about the reign of a woman as sign of the impeding end of time and connected them with descriptions of a sinking of Constantinople in the sea and a transfer of imperial power from Byzantium to Rome.[20] One of these texts, the so-called »Apocalypse of (Pseudo)Leon of Constantinople«, even refers to an »empress called Eirene« in this context.[21]

---

17  The Byzantine East mostly used a different calculation for the creation of the world, resulting in a date of c. 5500 BC (most commonly 5508 BC). Thus, the dawn of the 7th millennium of the world had been expected already for the time around AD 500, when a series of calamities in the reigns of the emperors Anastasius I (r. 491-518) and especially Justinian I (r. 527-565) were partly interpreted as apocalyptic omens; see Brandes, Anastasios ὁ δίκορος, 24-63; Meier, *Das andere Zeitalter Justinians*; Meier, The ›Justinianic Plague‹, 267-292; Preiser-Kapeller, *Der Lange Sommer*, 38-50 and 59-73. However, a number of apocalyptic texts from the 8th to the 9th centuries suggest an expectation of an end of times during these decades, at least in some circles of Byzantium; see Ubierna, L'apoca- lyptique byzantine, and Brandes, Traditions and expectations, 290-293, as well as below

18  Theophanes, *Chronographia*, vol. 1, ed. de Boor, 472, 18-22: ἐσκοτίσθη δὲ ὁ ἥλιος ἐπὶ ἡμέρας ιζ´ καὶ οὐκ ἔδωκε τὰς ἀκτῖνας αὐτοῦ, ὥστε πλανᾶσθαι τὰ πλοῖα καὶ φέρεσθαι, καὶ πάντας λέγειν καὶ ὁμολογεῖν, ὅτι διὰ τὴν τοῦ βασιλέως τύφλωσιν ὁ ἥλιος τὰς ἀκτῖνας ἀπέθετο. The use of the passive verb form ἐσκοτίσθη / eskotisthē (»was darkened«) indicates superior, i.e., divine will as the cause of the event.

19  *The Chronicle of Theophanes Confessor*, transl. Mango and Scott, 648-649; Kountoura Galaki, A light in the dark- ness, 143-158.

20  See the »Seventh Vision of Daniel«, dated to the late 5th century AD and preserved in an Armenian translation (La Porta, The Seventh Vision of Daniel, 428 und 431), and the »Diegesis of Daniel«, dated to the 8th or 9th centuries, Berger, *Die griechische Daniel-Diegese*, 14-15 and 91-92. See also DiTommaso, The Armenian Seventh Vision of Daniel; Berger, Das apokalyptische Konstantinopel; Ubierna, L'apocalyptique byzantine. For a most re- cent overview on apocalyptic texts in Byzantium, see Congourdeau, Textes apocalyptiques; Brandes, Traditions and expectations; and Kraft, An inventory of medieval Greek apocalyptical sources (also for the dating of the texts discussed above). On the interpretation of natural phenomena in these texts, see, most recently, Kraft, Natural di- sasters in medieval Greek apocalypses. On the general (ab)use of various forms of prognostication, also in political discourse, see Brandes, Kaiserprophetien und Hochverrat, and now Grünbart, Prognostication.

21  *L'apocalisse apocrifa di Leone di Costantinopoli*, cap. 15, ed. Maisano, 89-90. On the possible dating of the core of this text to the early 9th century (and its later adaptation in the 12th century), see Brandes, Sieben Hügel; Berger, Das apokalyptische Konstantinopel; Congourdeau, Textes apocalyptiques; Ubierna, L'apocalyptique byzantine; and now Kraft, An inventory of medieval Greek apocalyptical sources. Regarding the handling of apocalyptic expectations in the *Chronographia*, Torgerson, *Chronographia*, 37, observes: »Readers of the Chronographia found in the work not only an authoritative account of all past time but a definitive adjudication of time's end as the meaning behind the chronology of the Roman emperors. Accordingly, we find the Chronographia completely uninterested in anything resembling apocalyptic prophecies or calculations, and yet deeply invested in making meaning out of the time at hand through the figures of apocalyptic typologies.«

Interestingly, even the *Chronographia*, which otherwise tried to preserve a positive image of Eirene, linked Constantine VI´s blinding with the events leading to the imperial coronation of Charlemagne since immediately after the description of this event we read:

> In the same year, too, the relatives of the blessed Pope Adrian in Rome roused up the people and rebelled against Pope Leo, whom they arrested and blinded. They did not manage, however, to extinguish his sight altogether because those who were blinding him were merciful and took pity on him. He sought refuge with Karoulos, king of the Franks, who took bitter vengeance on his enemies and restored him to his throne, Rome, falling from that time onwards under the authority of the Franks. Repaying his debt to Karoulos, Leo crowned him emperor of the Romans in the church of the holy apostle Peter after anointing him with oil from head to foot and investing him with imperial robes and a crown on 25 December, indiction 9.[22]

This narrative links is also evident in the oldest manuscript tradition of the *Chronographia*.[23] Arguably, the chroniclers intended to contrast the brutal blinding of Constantine VI with the more »merciful« one of Pope Leo III; the two events, however, did not take place »*in the same year*« of AM 6289. Leo III was blinded on 25 April 799, that is AM 6291.[24] In order to bring the narrative on Leo III to a conclusion, the *Chronographia* further adds the reference to the crowning of Charlemagne, which took place on 25 December 800, that is AM 6293. This reference is later repeated in the *Chronographia* under the correct year AM 6293[25], which suggests that the authors were informed about the actual chronology of the events in Rome. Nevertheless, they put the blinding of Constantine VI and of Leo III  »*in the same year*« to serve their narrative strategy, culminating in the coronation of the Charlemagne as a con- sequence of the «darkening« of imperial power in Constantinople and papal power in Rome (somehow similar to the above-mentioned argument put forward by Alcuin).[26] Such adapta- tions of the chronology of events to the purposes of the narrative are quite common in the *Chronographia*,[27] also for celestial phenomena during the reign of Eirene and Constantine VI (see below).

---

A further narrative arc leading to the events of August 797 is already established earlier in the text: in August 792 (AM 6284), Constantine VI ordered the arrest, blinding and mutilation of his paternal uncles (after one of them, Nikephoros, had attempted a coup) and of some other high-ranking officials. According to the *Chronographia*:

> the punishment of those men took place in the month of August, on a Saturday, indiction 15, at the 9th hour. But not for long did God´s judgement leave this unjust deed unavenged: for after a lapse of five years, in the same month and also on a Saturday the same Constantine was blinded by his own mother.[28]

Via this parallelization, the otherwise horrible dimension of Eirene's deed, on the one hand emphasized through the comparison with the blinding of Leo III, is, on the other hand thus somehow reduced to a form of divine retaliation by the chroniclers.

For the darkening of the sun reported by the *Chronographia* for August 797, Paul Speck, in his highly erudite, but in parts poorly structured monograph on Constantine VI, supposed that this information originated from a »Vita of Constantine«.[29] The authors of the present article first thought that this must refer to Constantine the Great (r. 306/324-337)[30] and the famous *Vita Constantini* written by Eusebius.[31] Yet, there is no description of such a celestial sign in this text for the time of the death of the emperor. Rather, as can be learned from one (of a total of three) appendices to his work,[32] Speck had a Vita of Constantine VI in mind, which the *Chronographia* supposedly used as a source. Such a Vita, however, has neither been preserved nor is even secured in its existence, which Speck only postulated based on circumstantial evidence.[33]

Some scholars, such as Ralph-Johannes Lilie, suspected the description of the darkening to be a purely literary fiction. Indeed, historiography since antiquity had framed the fall or death of rulers with various portents.[34] Plutarch, for instance, wrote with regard to the murder of Julius Caesar in 44 BC:

> But the greatest of the divine miracles was the comet, which shone for seven days after Caesar's assassination and then disappeared again, and next to it the darkening of the sunlight. Throughout the year the disc of the sun rose pale and without brilliance, radiating but feeble warmth. The air remained cloudy and heavy because the warming rays were too weak to penetrate it, the fruit withered prematurely and fell half-ripe to the ground because of the cool weather.[35]

---

However, recent studies have demonstrated that the described atmospheric and climatic changes were caused by a volcanic eruption of Etna in 44 BC and the even more powerful Okmok eruption in Alaska in the following year.[36]

Equally, an actual natural background to such portents and phenomena mentioned in the *Chronographia* can usually be verified elsewhere. Already before this, the chroniclers had connected the beginning of the conflict between Constantine VI and his mother Eirene in Feb- ruary 790 (AM 6282) with a »terrible earthquake« in Constantinople.[37] When Constantine VI confined Eirene in the Palace of Eleutherios in October 790 (AM 6283), a fire damaged parts of the city center of the capital.[38] During the rebellion of the army corps of the Ar- meniacs (stationed in northeastern Asia Minor) against the emperor, another fire occurred in Constantinople on 25 December 792 after a thunderstorm.[39] After Constantine VI had separated from his first wife Maria and married Theodote in September 795, in the same AM 6288, in April (796), according to the *Chronographia*, a »terrible earthquake« occurred in Crete and then in May, another one in Constantinople, indicating divine disapproval.[40] An actual physical background can probably be expected to an even greater extent for the darkening of the sun associated with the blinding of Constantine VI in August 797, when the chroniclers, who a priori intended to draw of a positive image of the empress, had little rea- son to cast Eirene's act of violence in an even darker light through further fictitious omens.[41] Rather, the mentioned public discussion (»all said and agreed«) of the darkening of the sun suggests that the credibility of the historical work would have been damaged if this generally remembered celestial phenomenon remained unmentioned.[42]

For this physical background to the darkening reported for August 797, different assump- tions and approaches exist that are examined in the following pages and contrasted with more recent natural scientific findings. On this basis, we propose a more probable physical background to the celestial manifestation described by the *Chronographia* and interpreted by contemporaries in connection with the blinding of Constantine VI, which in turn – even in the narration of Georgos Synkellos and Theophanes – provided a prelude to the coronation of Charlemagne.

---

*The Darkening of 797: A Solar Eclipse?*

In the commentary to their widely used English translation of the *Chronographia*, Cyril Mango and Roger Scott explained the underlying event for the narrative on August 797 with two solar eclipses, which occurred in temporal proximity to the blinding of Constantine VI: »There was a total eclipse on 20 February 798 and another on 16 August. The latter would have been on the first anniversary of the emperor´s blinding.«[43] In the introduction to their book, they link their assumption of a solar eclipse with the interpretation of the event given by the *Chronographia*: »The blinding of Constantine VI is not excused: it is presented as cruel and wicked, a judgement that was evidently shared by God, who caused an eclipse of the sun to occur on that fateful day.«[44] However, a review of the course and the geographical coverage of the annular (i.e., »ring-shaped«) solar eclipse of 20 February 798 in the relevant NASA database shows that it was not visible at all in Constantinople, but only (partially) in Western Europe and on the Iberian Peninsula (*see Fig. 1*).[45] This also applies to the eclipse of 16 August 1798, which could only be observed in western and southern Africa and South America (*see Fig. 2*).[46] Another solar eclipse (that Mango and Scott overlooked) is dated 26 August 797 (the closest in time to the blinding of Constantine VI); however, this event too was invisible in Constantinople and could only be observed in America (*see Fig. 3*).[47] During this period, only one solar eclipse was at least partially visible in Constantinople (also overlooked by Mango and Scott) on 3 March 797 (*see Fig. 4*).[48] If one follows Mango and Scott's assumption that the *Chronographia* retrospectively linked a solar eclipse with the blinding of the emperor, the eclipse of 3 March 797 that preceded the events would fit. At the Bosporus, however, this eclipse was by no means accompanied by a complete, but only by a partial coverage of the sun.

---

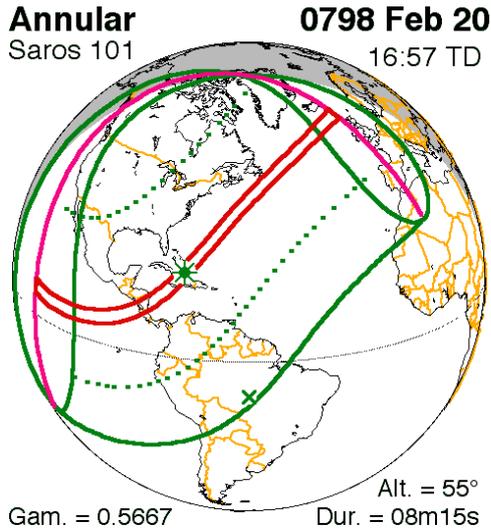

*Fig. 1: Umbra shadow (red) and zone of par- tial visibility (green) of the annular solar eclipse of 20 February 798 (Image source ac- cessed on 21 May 2021: eclipse.gsfc.nasa.gov/ 5MCSEmap/0701-0800/798-02-20.gif)*

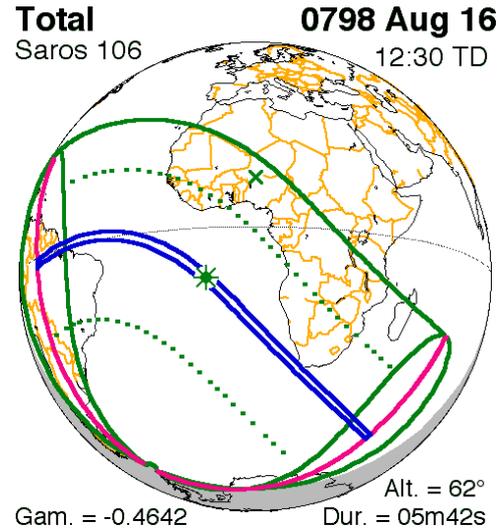

*Fig. 2: Umbra shadow (blue) and zone of partial visibility (green) of the total solar eclipse of 16 August 798 (Image source ac- cessed on 21 May 2021: eclipse.gsfc.nasa.gov/ 5MCSEmap/0701-0800/798-08-16.gif)*

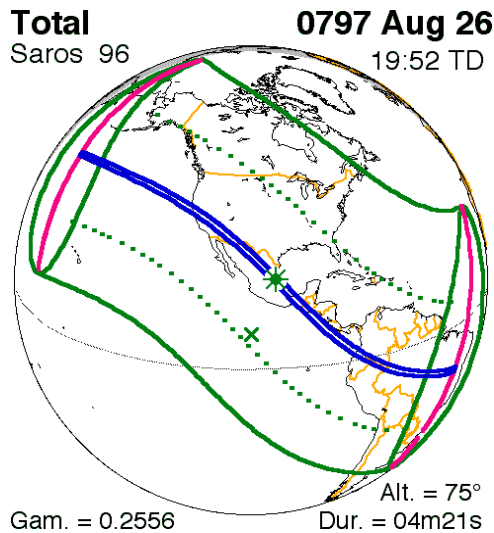

*Fig. 3: Umbra shadow (blue) and zone of partial visibility (green) of the total solar eclipse of 26 August 797 (Image source ac- cessed on 21 May 2021: eclipse.gsfc.nasa.gov/ 5MCSEmap/0701-0800/797-08-26.gif)*

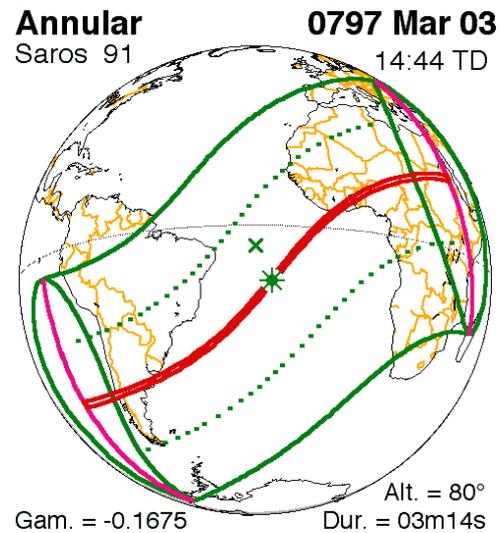

*Fig. 4: Umbra shadow (red) and zone of par- tial visibility (green) of the annular solar eclipse of 3 March 797 (Image source ac- cessed on 21 May 2021: eclipse.gsfc.nasa.gov/ 5MCSEmap/0701-0800/797-03-03.gif*

Furthermore (and quite importantly), if we take the *Chronographia*'s statement about the duration of this phenomenon seriously, it could not have been a solar eclipse at all! While the exact number of days for the darkening of the sun – 17 (ιζ΄) – may have been given in analogy to the 17 years of Constantine VI on the throne (September 780 un- til August 797, AM 6273-6289),[49] it nevertheless suggests a time period of several days or weeks. By contrast, the phase of maximum coverage of the sun can last up to approx- imately 7.5 minutes during a total solar eclipse, up to 12.5 minutes during a ring-shaped eclipse, and that of partial coverage, up to two hours, depending on the observation point.[50] Solar eclipses and their usual duration were also familiar and repeatedly observed phenomena for the Byzantines, and not likely to be mistaken for other phenomena.[51] Relatively close to the events of 797, the *Chronographia* describes a ἔκλειψις (eclipse) of the sun for the year 787:[52] »On 9 September of the 11th indiction, a Sunday, a considerable eclipse of the sun took place at the 5th hour of the day while holy liturgy was being performed.«[53] This was most probably the eclipse of 16 September 787; the core zone of the darkening migrated through the South Aegean, but it was also clearly visible in Constantinople (*see Fig. 5*).[54] The *Chronographia* mentions this eclipse for AM 6279 (1 September 786-31 August 787), al- though it took place in AM 6280 (1 September 787-31 August 788). But as Paul Speck has al-

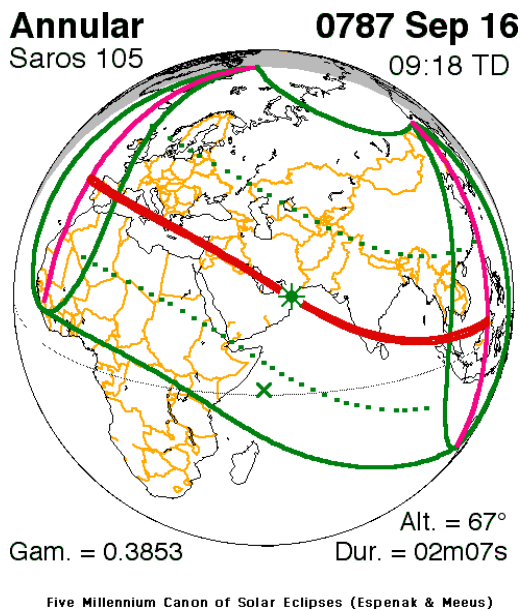

**Annular**       **0787 Sep 16**

Saros 105           09:18 TD

Gam. = 0.3853      Alt. = 67°     Dur. = 02m07s

Five Millennium Canon of Solar Eclipses (Espenak & Meeus)

ready suggested and the oldest manuscripts equally demonstrate,[55] the chroniclers re- served AM 6280 exclusively for the Second Council of Nicaea (which took place between 24 September and 23 October 787 and initi- ated a restoration of the veneration of holy images). They especially wanted to keep this year free from any portents which could have derogated the narrative of this (in their eyes) salutary event, such as a solar eclipse (of all things during a holy Sunday liturgy).[56]

*Fig. 5: Umbra shadow (red) and zone of par- tial visibility (green) of the annular solar eclipse of 16 September 787 (Image source ac- cessed on 21 May 2021: eclipse.gsfc.nasa.gov /5MCSEmap/0701-0800/787-09-16.gif)*

The *Chronographia* also provides relatively precise information on the duration of a much shorter-term event than that of 797 for 812:[57] »On 14 May, a Friday, there was a great eclipse of the sun lasting three and a half hours, from the 8th to the 11th hour.«[58] This time, the date given in the chronicle coincides exactly with the one based on modern calculations; the eclipse on 14 May 812 was visible in Constantinople as a partial eclipse (*see Fig. 6*).[59]

The last solar eclipse mentioned by the *Chronographia*, partially visible in Constantinople in the early morning, took place on 4 May 813 (see Fig. 7).[60] The exact time, celestial position (apparently determined by means of a »horoscope«) and psychological effect of the event are described in detail:[61] »On 4 May there was an eclipse of the sun in the 14th degree of the Bull according to the horoscope (κατὰ τὸν ὡροσκόπον[62]), at sunrise, and great fear fell on  the people.«[63] If this eclipse particularly worried the population of Constantinople, it was most probably due to the threat to the capital by the Bulgars under Khan Krum, which dominates the text passages in the *Chronographia* before and after the note about the celestial event.[64] Georgios Synkellos also makes a clear distinction between a short-term solar eclipse and a longer veiling of the sun in the first part of the *Chronographia* when he quotes from the discussion of the celestial phenomena at the crucifixion of Christ by Iulius Africanus:[65]

> In the third book of his Histories, Thallos [a historian of the 1st cent. CE] dismisses this darkness (σκότος) as a solar eclipse (ἔκλειψις). In my opinion, this is nonsense. For the Hebrews celebrate the Passover on Luna, and what happened to the Savior occurred one day before the Passover. But an eclipse of the sun takes place when the moon passes under the sun. The only time when this can happen is in the interval be- tween the first day of the new moon and the last day of the old moon, when they are in conjunction. How then could one believe an eclipse took place when the moon was almost in opposition to the sun?[66]

Therefore, the equation of the darkening of August 797 with a solar eclipse proposed by Mango and Scott is physically impossible and does not coincide with other descriptions of eclipses in the *Chronographia*. Another explanation of the phenomena described by the chroniclers must be sought.

---

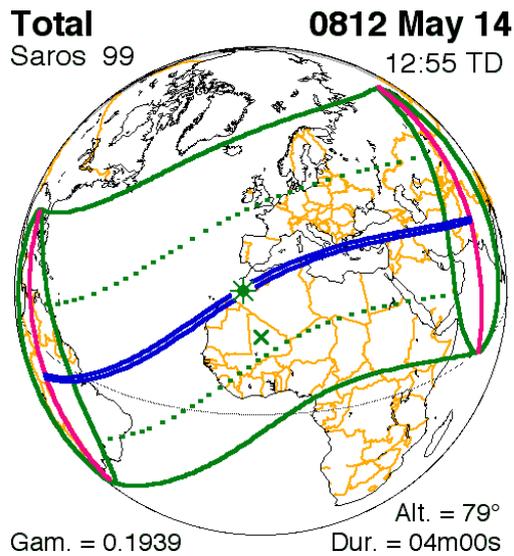

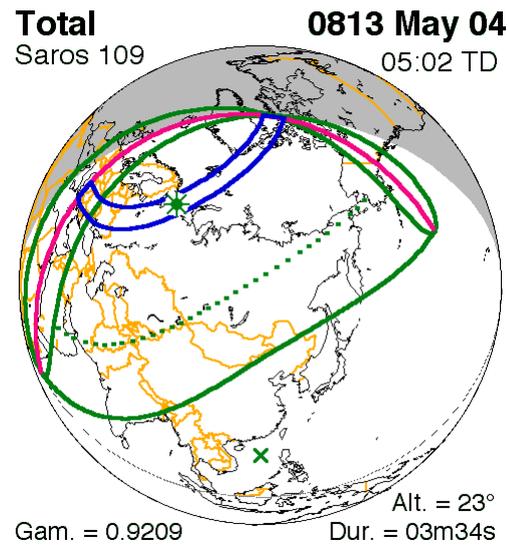

*Fig. 6: Umbra shadow (blue) and zone of partial visibility (green) of the total solar eclipse of 14 May 812 (Image source accessed on 21 May 2021: eclipse.gsfc.nasa.gov/ 5MCSEmap/0801-0900/812-05-14.gif)*

*Fig. 7: Umbra shadow (blue) and zone of partial visibility (green) of the total solar eclipse of 4 May 813 (Image source accessed on 21 May 2021: eclipse.gsfc.nasa.gov/ 5MCSEmap/0801-0900/813-05-04.gif)*

*Atmospheric Phenomena and Volcanic Eruptions Before and After AD 797*

Other scholars have also briefly discussed the prolonged darkness of August 797 both before and after Mango and Scott: Ioannis Telelis recorded the darkening in his catalog of meteorological phenomena in 2004, but without further consideration of an actual natural background. Rather, he assumed that the *Chronographia* juxtaposed the fading sunlight with the loss of the emperor's eyesight as a literary device. In her commentary, Ilse Rochow considered a longer period of bad weather as a background to the description of the *Chronographia*, since it would have been extremely unusual for this time of year (August). However, it is equally difficult to reconcile such a weather phenomenon with 17 days of darkening.[67]

As early as 1972, however, Robert R. Newton linked the darkening of August 797 with the atmospheric effects of a volcanic eruption, although he also considered that the description of the *Chronographia* was »probably a piece of propaganda directed against (Empress Eirene)«.[68] D. Justin Schove suspected a similar connection with a volcanic eruption in 1984, as did recently – following Newton – Thomas Wozniak in his impressive monograph on »Natural Events in the Early Middle Ages« in 2020.[69]

---

In fact, the material ejected from an explosive volcanic eruption can be hurled at different heights into the atmosphere and can contribute to an increase in atmospheric turbidity and create various optical phenomena – such as a weakening or scattering of sunlight. Due to patterns of global atmospheric circulation, over the course of a few days or weeks this can happen even at a great distance from the volcano. These phenomena are then visible for much longer than a solar eclipse; the »typical dwell times of the [volcanic] aerosols in the atmosphere depend on the particle size. In the lowest atmosphere they are days to weeks, in the stratosphere months to years.«[70]

At »17 days«, the »darkening« of the year 797 would have been of medium duration,[71] comparable to, but shorter than the famous »dust veil« of the year AD 536, described by Procopius, Cassiodorus, and other contemporary authors. Modern scholarship assumes a large explosive volcanic eruption, which left its traces in various proxy data such as ice cores and tree rings, as the cause for the 536 phenomenon. This eruption, together with another eruption in 540, initiated a cold anomaly, which may have provided the tipping point for a more prolonged colder climatic period – the so-called »Late Antique Little Ice Age« (LALIA), between 536 and 660.[72] Procopius wrote,

> And it came about during this year [AD 536] that a most dread portent took place. For the sun gave forth its light without brightness like the moon, during this whole year, and it seemed exceedingly like the sun in eclipse, for the beams it shed were not clear nor such as it is accustomed to shed. And from the time when this thing happened  men were free neither from war nor pestilence nor any other thing leading to death.[73]

---

70  Vollmer, *Atmosphärische Optik*, 66-67 (for the citation), 259-261, 279-281, 295-296. Cf. also Oppenheimer, *Erup-tions*, 53-76.

71  As one of the anonymous reviewers commented, »17 days would have been quite short for a large dust-veil (or aerosol haze, as it might also be known). However, it may be that the aerosol layer (which would be mainly sulfate) was at its greatest density at this location for 17 days, after which it may have still been present, but less notably in its impact on the visibility of the solar disk. Additionally, if a significant amount of the atmospheric material was instead composed of volcanic ash, this would last less time in the atmosphere, on the scale usually of only weeks (because the ash particles are often heavier than the aerosols). But if so, the volcano would probably need to have  been closer to the location of observation – perhaps Mediterranean or Icelandic [see also our deliberations on the  location of the volcano below, JPK and EK]. If it were only composed of sulfate aerosols, it could have been from  an eruption much further afield, in the tropics or northward.«

72  Gunn, *The Years without Summer*; Arjava, The mystery cloud of 536 CE, 73-94; Gräslund and Price, Twilight of the gods?, 428-443; Luterbacher and Pfister, The year without a summer, 246-248; Haldon *et al.*, Climate and environment of Byzantine Anatolia, esp. 123; Stathakopoulos, *Famine and Pestilence*, 265-268; Abbott *et al.*, What caused terrestrial dust loading, 421-437; Rigby *et al.*, A comet impact in AD 536; Büntgen *et al.*, Cooling and so- cietal change, 231-236; Newfield, Climate downturn of 536-50, 447-493; Newfield, Mysterious and mortiferous clouds, 89-115; Sarris, Climate and disease, 511-538; Büntgen *et al.*, Prominent role of volcanism; Preiser-Kapeller, *Der Lange Sommer*, 29-33 and 38-59.

73  Procopius, Vandalic War II 14, 5-6, ed. Haury and Wirth, vol. I, 482-483. English translation Dewing, *History of the Wars*, 329. See also Meier, *Das andere Zeitalter Justinians*, 363; Leppin, *Justinian*, 206-207.

The *Chronographia* provides a paraphrase of Procopius´ account (including the comparison with a solar eclipse, but making clear that it was a phenomenon of a different kind):[74]

> At this time, a portent occurred in the sky. For a whole year, the sun shone darkly, without rays, like the moon. Mostly it looked as if it was eclipsed, not shining clearly as was normal. It was the tenth year of Justinian's rule. In this time neither war nor death stopped weighing upon men.[75]

Ninety years after the spectacular dust veil of 536, Syriac sources from the northern Mesopotamian-Syrian region describe a similar phenomenon. For the year 626/627, we read in the *Chronicle* of Theophilos of Edessa, transmitted in the work of Michael the Syrian: »The light of one half of the orb of the sun was extinguished and there was a darkening from Oc- tober until June so that people said that the orb of the sun would never again be restored.«[76] As recent scientific research has demonstrated, this phenomenon can be connected to an- other volcanic eruption at that time, which, similarly to the 536 dust veil, contributed to a wide-ranging cold anomaly between the years 626 and 632 that was not only perceptible in the Mediterranean but also in the steppes of East Asia.[77]

Two further cases of anomalous atmospheric opacities are reported for the region of Syria and Northern Mesopotamia in the mid-8th century, first for the year 746.[78] In the chronicle of Michael the Syrian we read:

> In the same year [746], from the beginning of adar [March], until the middle of nisan [April], a kind of dust filled the whole atmosphere with darkness. All day long, the dust hovered in many places, and around nine o'clock it formed an opacity that hid the rays of the sun.[79]

---

For August 747, it is reported: »There was an intense darkness for five days in August. The atmosphere was turbid and opaque. The sun was like blood and its light weak. However, it was not an eclipse, but a turbidity of the atmosphere.«[80] Of interest is the overlap of this event with the last great outbreak of the first plague pandemic in Constantinople in 747/748, as also reported by the *Chronographia*. A further inquiry into possible climatic-epidemic connections is an attractive prospect, but beyond the scope of the present paper.[81]

Referring to such source evidence, Jonny McAneney likewise wrote in his online article »The mystery of the offset chronologies: Tree rings and the volcanic record of the 1st millennium« that the description of the *Chronographia* for 797 »is suggestive of a volcanic dust veil or ash cloud observed from Constantinople, possibly from a Mediterranean eruption«.[82] Richard B. Stothers, whilst concurring on a possible volcanic origin, assumed a volcanic eruption in Iceland.[83] In principle, this would be possible: after the eruption of the volcano Eyjafjallajökull took place on 14 April 2010, the ash cloud had spread to Istanbul by 18 April (*Fig. 8*).[84]

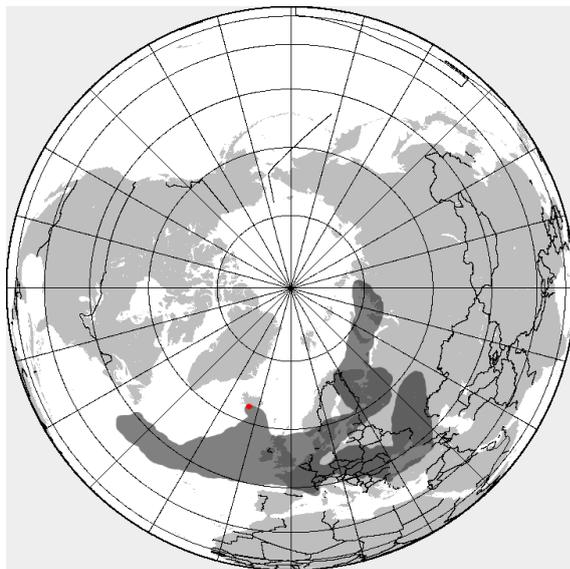

*Fig. 8: Expansion of the ash cloud after the eruptive eruption of Eyjafjalla- jökull in Iceland on 14 April 2010, until 18 April 2010 (Image source accessed on 21 May 2021: commons.wikimedia.org/wiki/File:Eyjafjallaj%C3%B6kull_volcanic_ash_18_April_2010.png*

Phenomena like those observed after the dust veil of 536, or in the year 797, were also reported after the eruptions of Laki in Iceland on 8 June 1783 and of Tambora in what is now Indonesia on 5 and 10 April 1815, respectively. In 1783, various con- temporary observers called the turbidity of the atmosphere caused by the volcanic aerosols »high-altitude smoke« (in German »Höhenrauch«). It spread along the main wind directions over thousands of kilometers from Iceland towards Europe, Asia and Africa and sometimes lasted from mid-June to August or September (thus even longer than the 17 days mentioned by Theophanes in 797). As Alan Mikhail demonstrated in an article about the effects of the 1783 eruption in the Ottoman Empire, the »Höhenrauch« was noticeable as far

---

as Constantinople and further south to Lebanon (*Fig. 9*).[85] During the Laki eruption, around 120 million tons of sulfur dioxide gas were injected into the atmosphere, so that eye and nose witnesses also reported a sulfur-like odor and problems with breathing. In addition, the atmospheric turbidity and particularly heavy fog formation hindered shipping in the North Atlantic and the Mediterranean (for example in the region around Malta) for several days – a phenomenon similar to that the *Chronographia* describes for 797.[86]

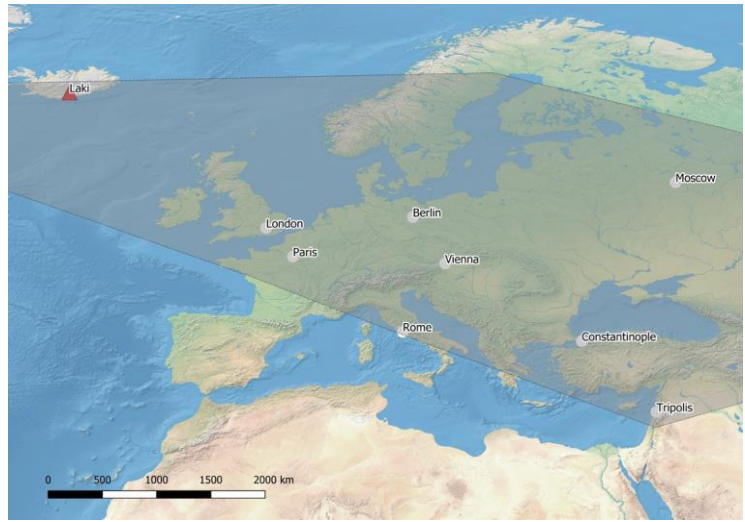

*Fig. 9: The expansion of the »Höhenrauch« (haze) after the eruption of the Laki volcano in Iceland on 8 June 1783 during the following summer according to eyewitnesses (adapted from: Mikhail, Ottoman Iceland; map: J. Preiser-Kapeller, 2022)*

*Observations of Atmospheric Opacity Beyond Constantinople in AD 797?*

The effects of the two volcanic events of 1783 and 1815 were perceived worldwide. However, what about similar observations to those made in Constantinople in the summer of 797 in other regions?

In his article of 2002, Stothers discussed the relevant passage in the *Chronographia* and added: »French chroniclers, some of them contemporary with Theophanes, and also later German chroniclers, have recorded that the *sidus Martis* (the southern constellation Scorpius) could not be seen from July 797 to the following July.«[87] In fact, there are entries that the *sidus, quod dicitur Martis* was not visible anywhere in the night sky from July 797

---

to July 798 (incidentally, immediately after the mention of an embassy from Empress Eirene to Charlemagne, on which occasion the deposition and blinding of Constantine VI are also reported)[88] in the *Annales regni Francorum* (also called *Annales Laurissenses maiores et Einhardi*)[89] and the *Annales Tiliani.*[90] The reference to this observation was also adopted in later sources, such as in the world chronicle of Abbot Regino of Prüm from around 967.[91]

Stothers thought that the star mentioned was Antares, also known as *sidus Martis* (»Mars star«), the brightest star in the constellation Scorpius and the 16th brightest star in the night sky; it received this name due to its red color, which is similar to that of planet Mars (as in the Greek Ἀντάρης as the »counterpart« to Ares/Mars). From Central Europe, the constellation Scorpius is only visible in the summer months and then only partially just above the southern horizon; haze layers near the horizon can interfere with the observation (*see Fig. 10*).[92] A general veiling of the atmosphere, as can occur after a major volcanic eruption, could have hindered the visibility of this otherwise conspicuous star.[93]

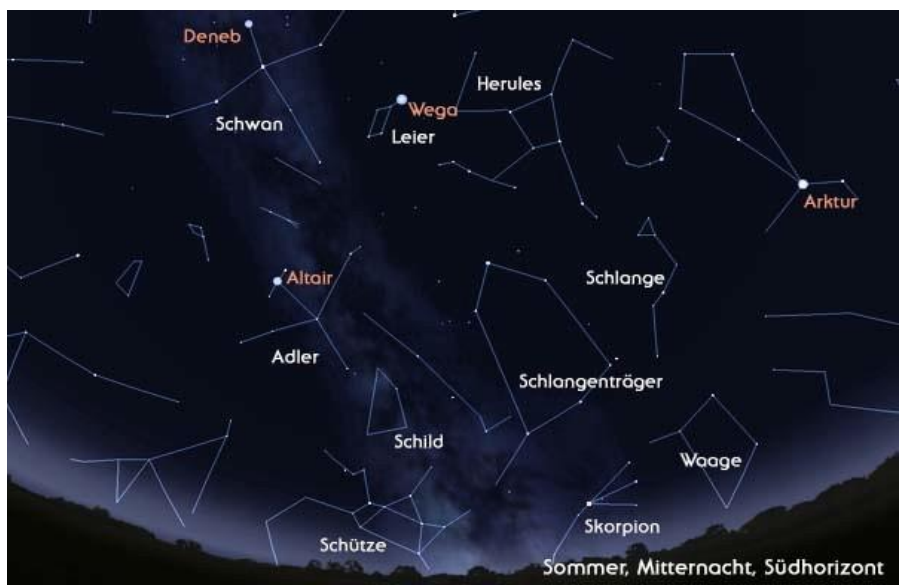

*Fig. 10: The southern starry sky in the summer months in Central Europe with the constellation Scorpius (in German »Skorpion«) and the star Antares just above the horizon (source accessed on 21 May 2021: astrokramkiste.de/himmel-im-sommer)*

---

However, Stothers' interpretation proves to be a mistake. In a letter dated September 798, the scholar Alcuin mentioned above, who had been part of Charlemagne's retinue since 781,[94] replied to a question from the king regarding the recently reappeared *stella Martis*, which could not be observed for a long time. From Alcuin's description of the various places in the firmament where this star was recently visible or »invisible«, it becomes clear that it was the

»wandering star« (planet) Mars (which is also explicitly stated in the text several times), and not the »fixed star« Antares.[95]

Charlemagne had evidently hoped that the reappearance of the planet of the god of war would be an omen of future success in warfare, an interpretation to which Alcuin only reacted evasively. As Kerstin Springsfeld explained in detail in her study, the temporary »disappearance« of Mars from the starry sky in certain latitudes, which included the Carolingian Empire, can be explained by the usual celestial mechanics (and is also retrospectively calculable for the relevant months 797/798), and does not demand the hypothesis of a turbidity of the atmosphere.[96]

### Ice Cores, Volcanic Eruptions, Cold Anomalies, and Their Dating

The Frankish sources cited by Stothers therefore do not appear useful as parallel reports for an atmospheric event at the time of the veiling of the sun in Constantinople in August 797. How- ever, he mentioned other indications for the hypothesis of a volcanic eruption as the cause of the events in summer 797: »This is supported by an ice core recovered at the Crete station on Greenland (although not by any others), in which the largest acidity peak between the years 626 and 934 falls in the year 798 ± 2 [i.e., between 796 and 800] (…).« [97] The idea of using ice cores from Greenland to corroborate historical reports of possible volcanic atmospheric anomalies, which – similar to the volcanic eruption responsible for the dust veil of 536 – would preserve the signature in deposited layers of past precipitation of the chemical agent responsible for anomalous atmospheric optical effects arising from past eruptions in or around 797, was sup- ported by McAneney in 2015 (see above with note 76).[98] He hypothetically linked them to the results of the analyses of ice cores of the North Greenland Eemian Ice Drilling project, but with a peak in sulfate values that he claims to be dated between 793 and 795 (and not between 796 and 800, as Stothers said), which would indicate a major volcanic eruption during this period.[99]

In an indirect way, reports of extreme weather conditions and crop failures between 792 and 794 from the Carolingian Empire may indicate that such an event could actually have taken place in the earlier 790s. Similar to the eruption of 536, it could have influenced the climate in the Northern Hemisphere chiefly through the atmospheric effect of sunlight-reflecting sulfate aerosols. Such a scenario receives additional support from tree ring data, which indicate significant impairment of plant growth from cold in these years (see Fig. 11 for an example from modern-day Switzerland).[100] As Reinhold Kaiser noted, Charlemagne reacted to the resulting first great famine of his reign, by setting maximum prices for bread and grain [at the Synod of Frankfurt in 794], introducing new measures, weights and coins, and imposing prayer services on the bishops' churches and monasteries as well as – like for the lay elites (counts, vassals) – taxes, depending on their possessions, for the feeding of the poor.[101]

The construction of a canal between the rivers Altmühl and Rezat (in Middle Franconia, the so-called »fossa Carolina«), which was started in 792/793 and was supposed to connect the river systems of the Danube and the Rhine, had to be abandoned by Charlemagne due to unfavorable weather conditions and probably also due the general distress in the empire.[102]

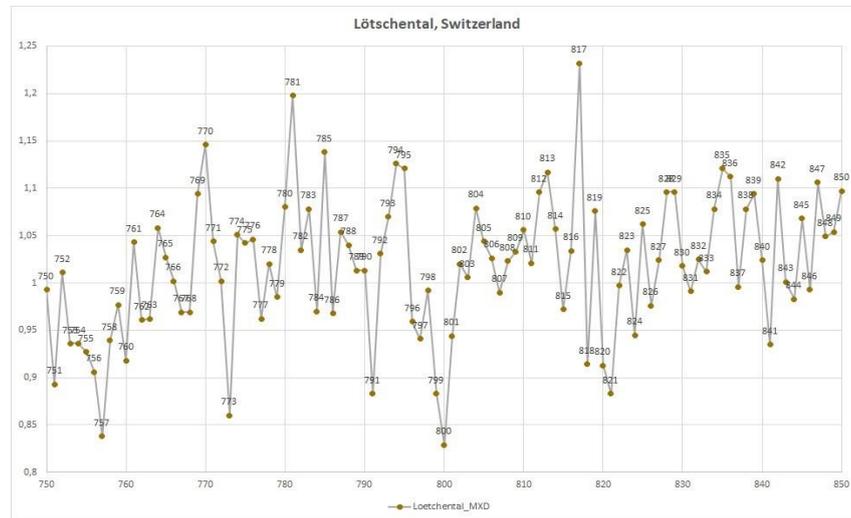

*Fig. 11: Tree ring data from the Lötschental in Switzerland for the period 750 to 850 (data: Guillet* et al.*, Climatic and societal impacts; graphic: J. Preiser-Kapeller, 2022)*

However, for the last decade of the 8th century, a refined dating of the ice core data from Greenland now confirms the occurrence of a major volcanic eruption only for the years around 800. An earlier eruption in the 790s instead finds little evidence in the newly dated Greenland ice core data (*see Fig. 12*). This does not exclude the possibility that an eruption with some climatic impact did happen, but with relatively little sulfate deposition occurring over Greenland (*see also below*).

---

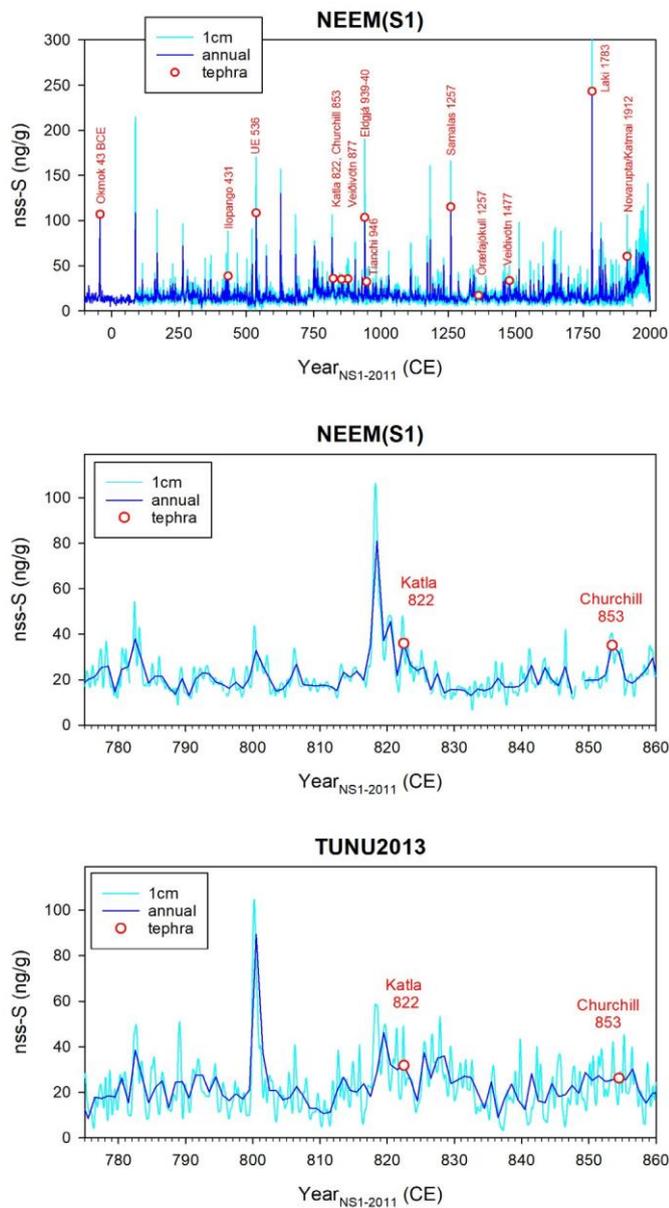

*Fig. 12: **(Top)**: NEEM(S1) Greenland ice core: non-sea-salt-sulfur concentrations on the NS1- 2011 chronology (constrained with 775 CE 10Be/14C anomaly).[103] Red circles are eruptions at- tributed on the basis of the geochemistry of tephra shards extracted from ice core records in Greenland. **(Middle)**: the time window 775-860 CE from the NEEM(S1) Greenland ice core data shown at the top. **(Bottom)**: the time window 775-860 CE from the TUNU2013 Greenland ice core record. The Katla 822/823 eruption is dated by dendrochronology[104] and tephra was identi- fied in the TUNU2013 ice core.[105] The Mt. Churchill eruption (White River Ash B) was identified by tephra in the NEEM and NGRIP ice cores[106] (Graphs by courtesy of Michael Sigl).*

---

The new dating of volcanic eruptions during these centuries, among other factors, made use of the discovery, published in 2012, that an off-planet event in 774/775 had dramatically enhanced the production of the radioactive carbon-14 isotope in the earth's atmosphere. This was detected first in the chemical signature of tree rings from Japan and then in many other datasets worldwide.[107] The cause of this phenomenon was probably a particularly strong outbreak of the sun (»flare«) or even a so-called coronal mass ejection, which sent a large number of charged particles towards the earth, where they interacted with atoms in the high Earth atmosphere, producing an excess of radioactive carbon-14, beryllium-10 and chlorine-36. While our current electronic and digital infrastructure would be very sensitive to such an event, most people in the 8th century would not have been materially affect- ed. However, sightings of auroras around this time not only in England, but also far south around Amida in northern Mesopotamia (as well as in China), identified by F. Richard Ste- phenson, Hisashi Hayakawa and others, also point to an enormously increased solar activity around the »774/775 event«.[108]

Most recently in 2022, a synthesis of the new chronology was published based on a syn- chronization of seven Greenland ice core chronologies (*see the map Fig. 16 for these sites*). As »tie points« between these datasets, the study uses chemical traces of significant volcanic eruptions (especially sulfates, SO4) and peaks in ammonium (NH4+) concentration, which can be found in all ice core sequences (*see Table 1*).[109] Ammonium peaks in the Greenland ice cores are usually the result of depositions from large scale wildfires, especially in adjacent Northern America (modern-day Canada), some of which may have thus taken place during the period under consideration in the present paper, especially in or around 796 (*see Table 1*). Potentially (as one of the anonymous reviewers also pointed out), particles released by major forest fires, and even by large-scale dust and sand transportation from places like Eurasia, North America or North Africa, can have similar atmospheric optical effects to those caused by volcanic ash and sulfates. However, on the basis of current knowledge, it is unclear if the source of the »major ammonium match« of c. 796 could have had any connection to atmos- pheric phenomena described during this period in Europe. Historically, volcanic aerosols have more usually been the culprits for prolonged or geographically widespread observa- tions of unusual atmospheric phenomena, including a veiling of the sun.[110]

---

| Year AD | Season of signal | Type of signal | Years before 2000 CE | Un-certainty margin of dating, in years | Range of dating |
|---|---|---|---|---|---|
| 750 | July-September | **Major volcanic match** | 1249.36 | 4.25 | 746-754 |
| 758 | July-September | Minor ammonium match | 1241.49 | 4.23 | 754-762 |
| 762 | July-September | **Major volcanic match** | 1237.34 | 4.22 | 758-766 |
| 785 | April-June | **Major volcanic match** | 1214.52 | 4.17 | 781-789 |
| 792 | April-June | Minor ammonium match | 1207.62 | 4.15 | 788-796 |
| 793 | July-September | Minor ammonium match | 1206.40 | 4.15 | 789-797 |
| 796 | July-September | **Major ammonium match** | 1203.29 | 4.15 | 792-800 |
| 798 | July-September | Minor ammonium match | 1201.45 | 4.14 | 794-802 |
| 799 | April-June | **Major volcanic match** | 1200.63 | 4.14 | 795-803 |
| 805 | July-September | Major ammonium match | 1194.392 | 4.12 | 801-809 |
| 810 | July-September | Major ammonium match | 1189.29 | 4.11 | 806-814 |
| 822 | April-June | **Major volcanic match** | 1177.55 | 4.08 | 818-826 |
| 824 | July-September | Minor ammonium match | 1175.28 | 4.08 | 820-828 |
| 831 | July-September | Minor ammonium match | 1168.46 | 4.06 | 827-835 |
| 832 | April-June | Minor ammonium match | 1167.52 | 4.06 | 828-836 |
| 836 | July-September | Minor ammonium match | 1163.41 | 4.05 | 832-840 |
| 843 | July-September | Minor volcanic match | 1156.37 | 4.03 | 839-847 |
| 845 | April-June | Major ammonium match | 1154.51 | 4.03 | 841-849 |
| 846 | July-September | Minor volcanic match | 1153.32 | 4.03 | 842-850 |
| 847 | July-September | Minor ammonium match | 1152.33 | 4.02 | 843-851 |

*Table 1: Major and minor volcanic events and ammonium peaks in a new synchronized chronolo- gy based on seven Greenland ice cores for the mid-8th to mid-9th centuries CE (data from: Sinnl et al., A multi-ice-core, annual-layer-counted Greenland ice-core chronology).*[111]

These revised chronologies also contribute to the disproving of another hypothesis on the volcanic source of origin for the »darkening of the sun« of 797 proposed by the geologist Johannes Koch in 2010: Mount Churchill in Alaska. Koch stated:

---

111 As one of the anonymous reviewers pointed out, the table does not necessarily provide a complete record of rel-evant volcanic signals in the polar ice, since it shows only the events thought to match between the different ice cores. The record in Sigl *et al.*, Timing and climate forcing, for instance also shows a major event in 817, which seems to have only deposited sulfate in Greenland and not also in Antarctica. In general, ice cores can be dated by counting annual layers in their upmost layers. Otherwise, common stratigraphic markers (paleo-events with a secure dating and a clearly identifiable chemical signature) can be used, such as the »774/775 event«. In addition to geochemistry, other methods such as the measurement of electrical conductivity may be applied. All these methods together usually still leave a margin of uncertainty in dating also indicated in Table 1.

Here, we report an estimated calendar age for the White River Ash from Mount Churchill, Alaska, one of the largest eruptions in the past two thousand years, using Bayesian statistics. Excavation of trees killed by the eruption, tree-ring cross-dating, and radiocarbon ages provide a new age range for the eruption of 750 to 813 AD (95% credible interval), with a modal value of 796 AD. These results suggest that the anomalous atmospheric event of 797 AD was caused by the Mount Churchill eruption.[112]

The Mount Churchill eruption mentioned by Koch was certainly enormous; ejected material can be detected not only in large parts of the North American continent, but as far away from Alaska as Central Europe.[113] As mentioned above, a 2020 study also showed that an eruption of the Okmok volcano on one of the Aleutian Islands off Alaska in the spring of 43 BC was capable of inducing atmospheric and climatic effects in the Mediterranean area, linked by ancient authors with the murder of Julius Caesar in 44 BC.[114] However, all recent studies on the eruption of Mount Churchill assume a date for the eruption between 833 and 853 (the later date is now considered to be fairly certain, *see Fig. 12*); thus, the assumption made by Koch about the connection with the darkening of the sun in Constantinople in 797 loses its chronological basis.[115]

By contrast, the new chronology also identifies a »major volcanic match« in 762/763 (*see Table 1*), already described as a potential cause for a cold anomaly in winter 763/764 by Michael McCormick, Paul Edward Dutton, and Paul A. Mayewski in 2007, based on a combination of the data from Greenland available then and written sources. This anomaly is reported not only for Ireland and the Frankish Empire, but also for Constantinople and the Black Sea region by Georgios Synkellos and/or Theophanes, who as young boys became eyewitnesses of a freezing of the Bosporus and other sea regions around the capital.[116] The *Chronographia* also describes the psychological effect of this event in Constantinople:

All the inhabitants of the City, men, women, and children, ceaselessly watched these things and would return home with lamentation and tears, not knowing what to say. In the same year, in the month of March [764 CE], the stars suddenly fell from the sky, so that all observers thought it was the end of the present world age.[117]

---

Jesse W. Torgerson further elaborates on the possible interpretations of these portents:

> (…) the freezing of the coast of the Black Sea is described as meaning »the sea became indistinguishable from land,« in other words an undoing of the work of the third day of creation. (…) The falling of the stars is an undoing of the work of the fourth day [when God created the sun, the moon and the stars] – in traditional chronological thinking the day that marked the beginning of time itself.[118]

A similar interpretation regarding the undoing of the marking of the flow of time could be imagined in face of the darkening of the sun connected with the blinding of Constantine VI, when ships also lost their orientation due to the invisibility of the celestial bodies.

Stothers also assumed that the volcanic eruption he dated between 796 and 800 would have been powerful enough »to cause climatic cooling. The Annals of Ulster (A.D. 798) men- tion a ›great snow‹ in 798, while northern tree ring data indicate very cool summers in the period 794-800 (...).«[119] In fact, the Irish Annals of Ulster record, albeit under the year 799:

»A great fall of snow, in which many men and cattle died.«[120] In contrast to the anomaly of 763, which is also recorded in Irish chronicles,[121] however, there are no corresponding re- ports for the years 797/798 from other regions of Europe. Only the *Annales Flaviniacenses* from the French monastery of Flavigny for the year 797 offer a unique single entry: *Siccata fluminum idem maris*. Such a »drying up of the rivers as well as of the sea« is not mentioned in any other Frankish source of the time, so that it remains questionable whether such an ex- treme weather event actually took place.[122] To sum up, such indirect evidence from Western and Central Europe for 797/798 is scant, and not every regional climatic anomaly has to be explained by volcanic activity elsewhere.

There remains Stothers' statement that northern European tree ring data show very cold summers between 794 and 800.[123] Jonny McAneney also writes, as already mentioned, that it is »tempting to link this event [i.e. the darkening of the sun of 797] with the sudden cooling observed in Swedish pine that occurred in AD 800«.[124] The tree ring data from Scandina- via was recently re-evaluated and published together with data from the Alps by Sébastien Guillet and his team. The data series from Northern Europe (Torneträsk in Sweden *[Fig. 13]* and an analysis of several tree ring series from Northern Sweden and Finland *[Fig. 14]*) as well as from the Lötschental (*Fig. 11*) in Switzerland certainly show a marked low point in tree growth for the year 800, even more dramatic than for the early 790s (*see the map Fig. 16 for these sites*).

---

118 Torgerson, *Chronographia*, 225 (see also 116 on the significance of the fourth day of creation for the measurement of time, as well as Lempire, Traditions and practices).

119 Stothers, Cloudy and clear stratospheres.

120 *The Chronicle of Ireland*, transl. Charles-Edwards, 260 (and 35-39 on problems in the chronology of the Irish Annals). See also Ludlow *et al.*, Medieval Irish chronicles reveal persistent volcanic forcing.

121 *The Chronicle of Ireland*, transl. Charles-Edwards, 231.

122 *Annales Flaviniacenses*, ed. Pertz, 151. See also Wozniak, *Naturereignisse im frühen Mittelalter*, 526.

123 Stothers, Cloudy and clear stratospheres.

124 McAneney, Mystery of the offset chronologies.

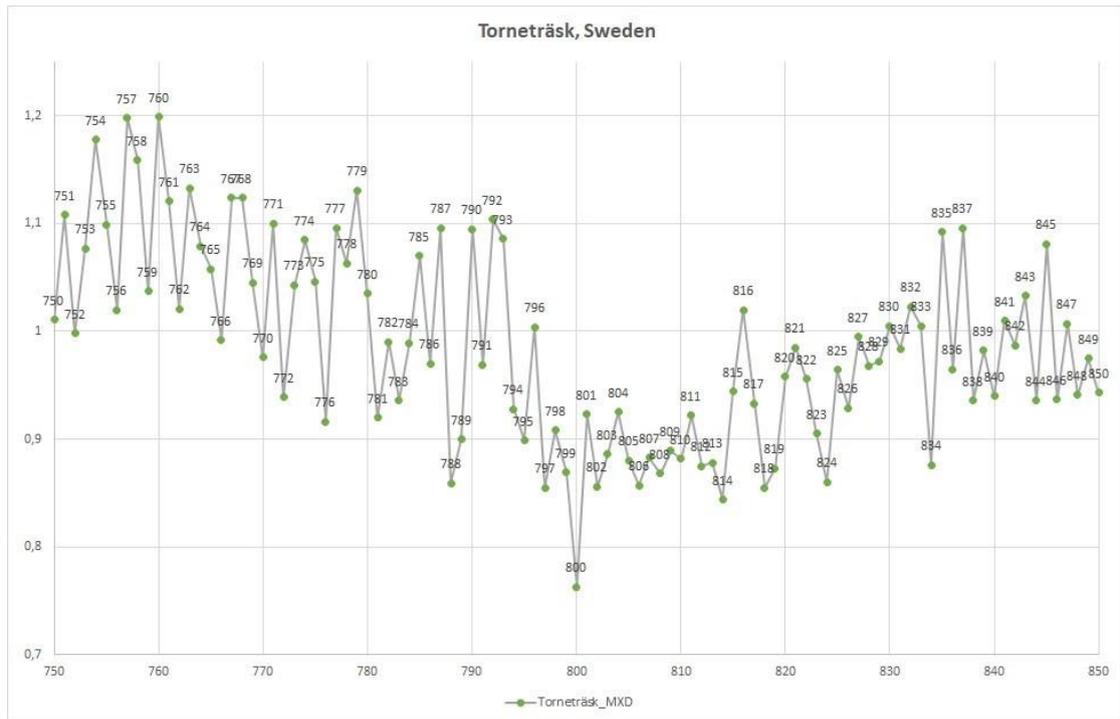

*Fig. 13: Tree ring data from Torneträsk in Sweden for the period 750 to 850 (data: Guillet* et al.*, Climatic and societal impacts; graphic: J. Preiser-Kapeller, 2022)*

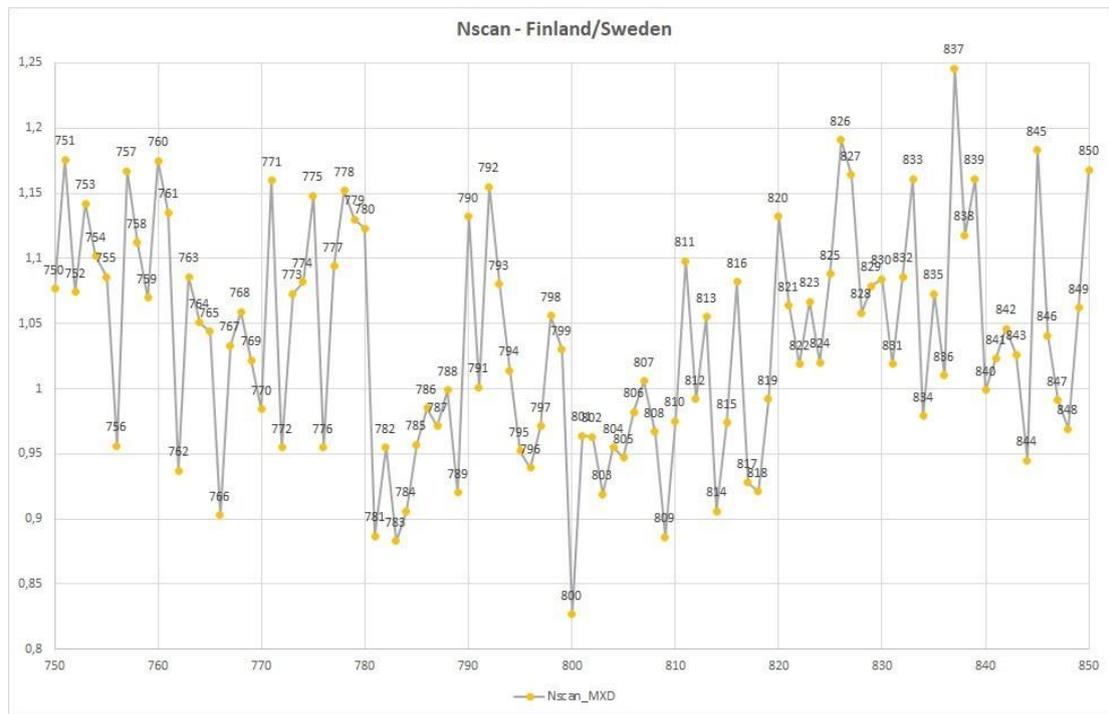

*Fig. 14: Tree ring data from northern Sweden and Finland for the period 750 to 850 (data: Guillet* et al.*, Climatic and societal impacts; graphic: J. Preiser-Kapeller, 2022)*

In the overall reconstruction of summer temperatures in the Northern Hemisphere by Guillet *et alii* for the last 1500 years, based on these data, the year 800 marks one of the lowest values ever, which is comparable with well-known volcanic cold anomalies such as in 536 or 1816 (after the Tambora eruption of 1815, the »year without a summer«) (*Fig. 15*).[125] Furthermore, the revised Greenland chronologies date a »major volcanic match« to c. 800 (*see Table 1*), without identifying the place of the eruption. Since the signal in Greenland finds no counterpart in data from Antarctica, possible candidates for this event would be volcanoes in the Northern Hemisphere, such as in Iceland, which were extremely active in the 8th to 10th centuries. Recently, a major eruption of Mt. Katla has been dated to 822 CE in connection with the new ice core chronologies from Greenland (*see Fig. 12*). Geologists likewise date other major eruptions of Mt. Katla or Mt. Hekla to around the year 800, but so far with a chronological margin (of up to several decades) that does not (yet) allow a more precise dating assignment in or at least close to the year 800.[126]

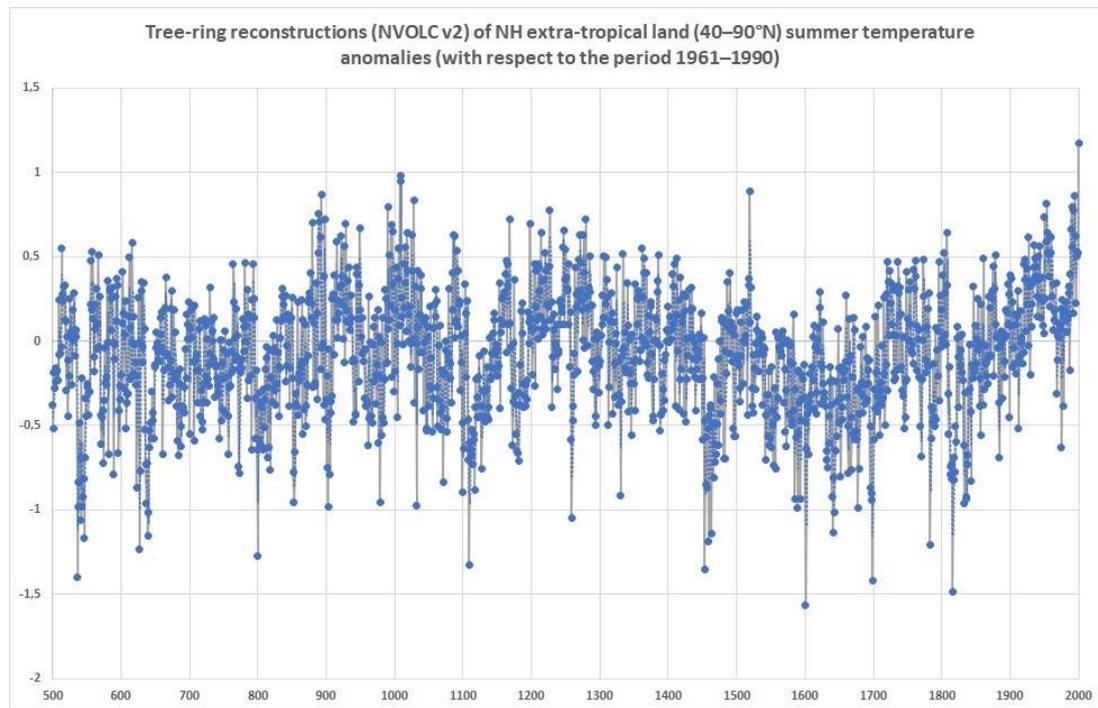

*Fig. 15: Reconstruction of summer temperatures in the Northern Hemisphere for the period 500 to 2003 (compared to the period 1961-1990) based on tree ring data from different regions (data: Guillet* et al.*, Climatic and societal impacts; graphic: J. Preiser-Kapeller, 2022)*

---

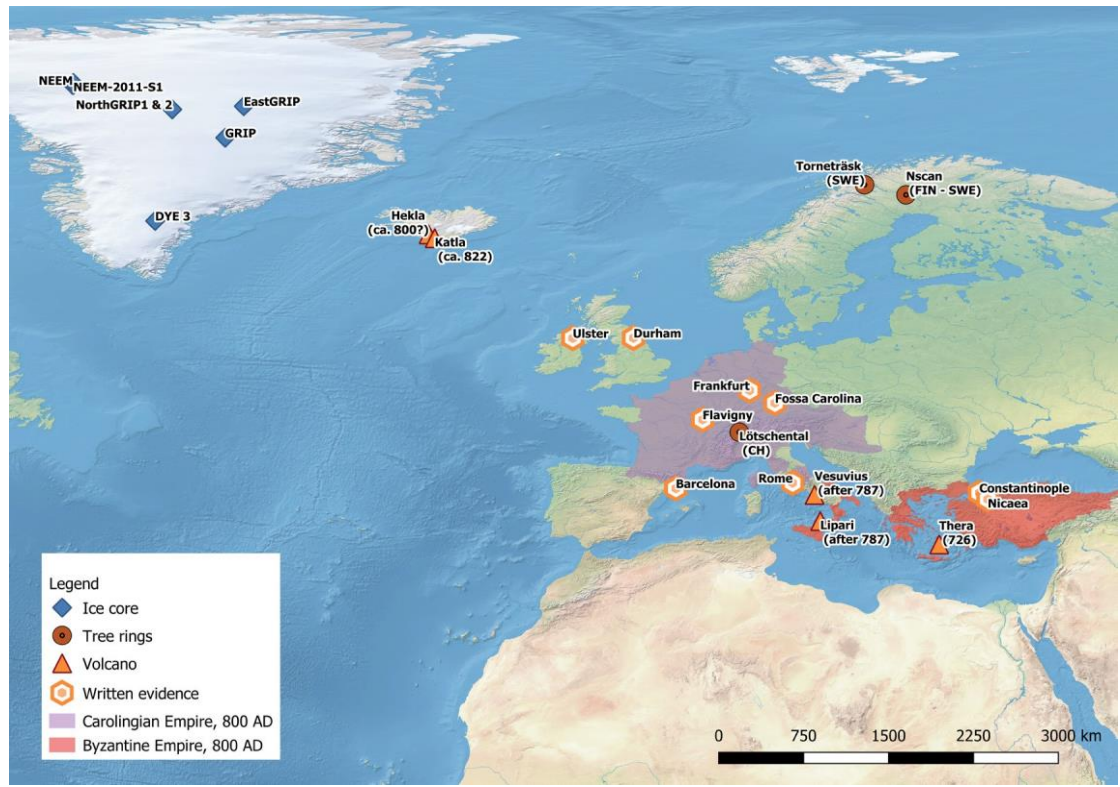

*Fig. 16: Map of selected historical and natural scientific evidence discussed in the paper for the late 8th and early 9th centuries CE (map: J. Preiser-Kapeller, 2022)*

Yet, while a weather anomaly in the year 800 left its mark on the tree rings in the high altitudes of the Alps and in northern Scandinavia, we do not hear anything about corresponding effects on arable crops. In the *Annales regni Francorum*, however, there is an interesting remark about June 800: »On 6 June and likewise on 9 June there was a severe frost which did not, however, harm the harvest.«[127] For the time around Christmas 800 (when Charlemagne was crowned emperor in Rome) sources for England (Simeon of Durham) report a storm and a storm surge that caused great devastation, and this was followed by a cattle epidemic.[128] The *Annales regni Francorum* for the year 801 also report a pestilence among animals and humans in the region of northern France and the Rhine. The reason given, though, is the
»gentleness« of the previous winter (*propter mollitiem hiberni*); also mentioned are »earthquakes in Germania and in Gaul in the places on the Rhine«.[129] Thus, a cold anomaly in summer 800 was perhaps also noticeable in the lower or more southerly located and more densely populated regions of Western Europe, but apparently had insufficient impact on agriculture for a famine to result from it, which would have found a more far-reaching echo in the sources.

---

127 *Annales regni Francorum*, ed. Kurze, 110 (*[P]ridie Non. Iul. insolito more aspera pruina erat et VII. Id. Iul. similiter, quae tamen nihil incommoditatis fructibus attulit*). Newfield, *Contours of Disease and Hunger*, 300, 428-429.

128 *Symeonis Monachi Opera Omnia*, ed. Arnold, 63; Whitelock, *English Historical Documents*, 250; Wozniak, *Natur- ereignisse im frühen Mittelalter*, 389, 454-455.

129 *Annales regni Francorum*, ed. Kurze, 114; Newfield, *Contours of Disease and Hunger*, 170-172, 188-189, 316, 428-429; Wozniak, *Naturereignisse im frühen Mittelalter*, 283, 516, 648, 674.

On the other hand, the biography of Louis the Pious, son of Charlemagne, states that during his siege of the Arab-ruled city of Barcelona in the winter of 800/801 the weather was so cold (*hiemis asperitate*) that the besieged hoped that the army of the Franks would be forced to withdraw. According to current climate data, even in the coldest month of January, the temperature in Barcelona hardly falls below plus 5 degrees Celsius. The Franks, however, built wooden shelters, maintained the siege, and thus forced the city to surrender in April 801 despite the unusual cold.[130]

An all too mild winter in northern France and on the Rhine, which nevertheless was harmful to human and animal health, at the same time as unusual winter cold around Barcelona and extreme cold in the high alpine locations and in northern Scandinavia (according to the tree ring data for the spring/summer 800), seems to create a contradictory scenario for 800/801 at first glance. However, it could point to the regionally different effects of a volcanic climate anomaly. As studies after more recent eruptions such as that of Pinatubo in 1991 have shown, depending on their quantity and distribution, volcanic aerosols result in patterns of atmospheric airflow, mediated by respective regional conditions, that may promote mild winters as well as the more commonly discussed cold winters.[131] In addition, volcanic eruptions can influence precipitation in various ways, most directly by reducing evaporation over water surfaces by cooling surface temperatures. This can reduce net pre- cipitation and river flow on various spatial scales, from regional to hemispherical to global, although because of changes in atmospheric circulation and airflow (as described above), regions might experience relatively wetter conditions.[132] For example, an extreme drought  has been reconstructed for the years from 800 onwards for Eastern Central Asia, to which explosive volcanism may have contributed, affecting the Empire of the Uyghurs (with its capital Ordu-Baliq) in modern-day Mongolia.[133]

For Byzantium, the *Chronographia* mentions that at the end of October 802, when Emperor Nikephoros I replaced Eirene on the throne in Constantinople:

> even the weather, contrary to nature, suddenly on that day became gloomy and light-less, filled with implacable cold in the autumnal season, clearly signifying the man´s future surliness and unbearable oppression, especially towards those who had chosen him.[134]

---

130  *Thegan - Astronomus*, ed. Tremp, 316, esp. 16-20; Wozniak, *Naturereignisse im frühen Mittelalter*, 744.

131  As one of the anonymous reviewers pointed out, »for northern and northwestern Europe, this is thought to opera- te through an invigorated westerly airflow, sweeping milder oceanic air in from the Atlantic, see Robock, Volcanic eruptions and climate, 191-219. This is generally thought to arise in the first winter after tropical (i.e., low latitude) eruptions. The bipolar ice-core data presented by Sigl *et al.*, Timing and climate forcing, suggests, however, that the 799/800 eruption occurred in the extra-tropical regions of the Northern Hemisphere, which may be less ex- pected to induce such a winter warming. But much depends upon the exact trajectory of the volcanic aerosol cloud resulting from this eruption.«

132  Stenchikov, Role of volcanic activity, 419-447; Ludlow and Manning, Revolts under the Ptolemies, esp. 165. See  also www.usgs.gov/natural-hazards/volcano-hazards/volcanoes-can-affect-climate (accessed on 21 May 2021),  and Iles and Hegerl, Systematic change in global patterns of streamflow.

133  Di Cosmo *et al.*, Environmental stress and steppe nomads, 439-463.

134  Theophanes, *Chronographia*, vol 1, ed. de Boor, 477, 14-18; *The Chronicle of Theophanes Confessor*, transl. Mango and Scott, 655-656. Telelis, *Meteorologika phainomena*, 365 (no. 287). On the interpretation of weather phenome- na see also Telelis, Traditions and practices.

While this passage provides an interesting parallel for how the *Chronographia* uses atmospheric phenomena to frame political events (clearly expressing antipathy towards the new emperor, which becomes evident from other parts of the chronicle), for the moment it does not allow us to reconstruct a more general cold anomaly in Constantinople at that time.[135]

Nevertheless, a climate-effective volcanic eruption can be assumed for the year 800, which is also evident in the newly dated Greenland ice core data (s*ee Fig. 12 and Table 1*). The darkening of the sun described by the *Chronographia* for Constantinople for August 797 could provide an indication of this eruption´s atmospheric-optical effect. Due to the practice by historians in the Middle Ages of combining significant natural events chronolog- ically with important political events such as the death or fall of a ruler and using them as a dramatic-symbolic background, we can hypothesize that corresponding celestial phenomena could have occurred sometime before or after the blinding of Constantine VI. in 797. In the later narrative of the *Chronographia* (and in the general »popular memory«, which is also echoed in the passage *Chronographia*) they were intertwined with the fall of the emper- or.[136] We have mentioned several examples where the *Chronographia* adapted the chronology of events to narrative needs (*see above*), and Mango and Scott put forward similar arguments with regard to the solar eclipses that took place several months after August 797, which they wanted to identify as background for the *Chronographia*´s description (*see above*).[137] Equally, McAneney wrote, »it is conceivable that the obscuration event could have been as much as a few years after Constantine's capture and blinding, but that the two events were associ- ated as direct cause and effect by the popular psyche at the time, and recorded as such.«[138] Thus, despite the stated date of the darkening of the sun over the Bosporus being nominally too early to be associated with the eruption dated on revised ice core chronologies to the  years 799/800, it is legitimate to consider them as potentially associated, in particular when also considering the small remaining uncertainties in these ice core chronologies (*e.g., Table 1*). It was only in retrospect that the *Chronographia* linked it with the blinding of Emperor Constantine VI in August 797 – as it linked other later events such as the blinding of Pope Leo III in 799 or Charlemagne´s coronation in 800 to Constantine´s downfall in the narrative for AM 6289 (*see above*).

---

### A Mediterranean Eruption in the Late 8th Century? Mt. Vesuvius, Lipari and Ashes at the Bosporus

Alternatively, it would be possible that a volcanic eruption happened as early as or before the summer of 797, which, like several eruptions well reconstructed from other evidence, did not make its way into the record in the ice cores in Greenland (or in the written evidence for Western Europe) due to atmospheric conditions, but caused the »darkening« in Constantinople. In his above-mentioned study, McAneney speculated that the description of the *Chronographia* for 797 »is suggestive of a volcanic dust veil or ash cloud observed from Constantinople, possibly from a Mediterranean eruption«.[139]

As the ice core scientist Michael Sigl, one of the leading researchers on historical volcanic eruptions, in discussion with us about our collected and examined findings remarked, and  as recent studies also demonstrated, several major eruptions in the Mediterranean did not leave clear chemical traces in the Greenland ice core records; this is even true for the most notorious and massive Mt. Vesuvius eruption of 79 CE, which destroyed Pompeii.[140]

The inhabitants of Constantinople, however, were generally quite familiar with the long-range effects of eruptions in the Mediterranean. About a »near-eruption« of Mt. Vesuvius in 536 and an earlier eruption, which can be dated to the year 472, Procopius reports the following:

> At that time [in the year 536] the mountain of Vesuvius rumbled, and though it did not break forth in eruption, still because of the rumbling it led people to expect with great certainty that there would be an eruption. And for this reason, it came to pass that the inhabitants fell into great terror. Now this mountain is seventy stades distant from Naples and lies to the north of it – an exceedingly steep mountain, whose lower parts spread out wide on all sides, while its upper portion is precipitous and exceedingly difficult of ascent. But on the summit of Vesuvius and about the center of it appears a cavern of such depth that one would judge that it extends all the way to the bottom of the mountain. And so, it is possible to see fire there, if one should dare to peer over the edge, and although the flames as a rule merely twist and turn upon one another, occasioning no trouble to the inhabitants of that region, yet when the mountain gives forth  a rumbling sound which resembles bellowing, it usually sends up not long afterward a great quantity of ashes. And if anyone travelling on the road is caught by this terri- ble shower, he cannot possibly survive, and if it falls upon houses, they too fall under the weight of the great quantity of ashes. But whenever it happens that a strong wind

---

comes on, the ashes rise to a great height, so that they are no longer visible to the eye, and are borne wherever the wind which drives them goes, falling on lands exceedingly far away. And once, they fell in Byzantium [Constantinople] and so terrified the people there, that from that time up to the present the whole city has seen fit to propitiate God with prayers every year; and at another time they fell on Tripolis in Libya. Formerly this rumbling took place, they say, once every hundred years or even more, but in later times it has happened much more frequently.[141]

A source for Procopius could have been Marcellinus *comes*, whose Latin chronicle extends to the year 534.[142] For the year 472, there is the following entry:

Under the consuls Marcianus and Festus, Vesuvius, the burned mountain of Campania boiling by inner fires, spewed burned bowels obscuring the light of day and covered all the surface of Europe with fine dust. This terrible dust is remembered every year at Byzantium, the 6th day of November.[143]

A little later, Ioannes Malalas and the *Chronicon paschale* (Easter Chronicle) likewise document this event.[144] The memory of the ash rain of 472 in Constantinople was preserved through the annual processions mentioned by Marcellinus Comes and Procopius.

Another possible episode of ashfall over the Bosporus from an eruption of Mt. Vesuvius is recorded in the apocalyptically inspired history of Leo the Deacon (c. 950-1000), who registered several disasters and portents for the year 967/968: he mentioned an earthquake in northwestern Asia Minor in 967 and a severe storm and flooding in Constantinople and its environs in June 968, so that:

---

141  Procopius, Gothic War II 4, 21-28, ed. Haury and Wirth II, 168-169. English translation Dewing, *History of the Wars*, 325, 327. For 472, see Grey, Risk and vulnerability, esp. 26, 28; Cioni *et al.*, Explosive activity and eruption scenarios, esp. 337, 341-343 (type subplinian I); Rolandi *et al.*, The A.D. 472 eruption of the Somma volcano, 291-319; Mastrolorenzo *et al.*, The 472 AD Pollena eruption of Somma-Vesuvius, 19-36; Stothers and Rampino, Volcanic eruptions in the Mediterranean before A. D. 630, 6361-6362; Rosi and Santacroce, The A.D. 472 ›Pollena‹ eruption, 249-271; Colucci Pescatori, Osservazioni su Abellinum tardoantica, 121-141; Grattan, Aspects of Arma- geddon, 13-14; Wozniak, *Naturereignisse im frühen Mittelalter*, 324, 348. For the procession on the anniversary of the event of 472, see Meier, *Das andere Zeitalter Justinians*, 493-494. On an eruption of Vesuvius in 512, see Cioni *et al.*, The 512 AD eruption of Vesuvius, 789-810; Kostick and Ludlow, Dating of volcanic events.

142  Croke, *Count Marcellinus and his Chronicle*.

143  *Marcellini v. c. Comitis Chronicon, ed. Mommsen, 90: Vesuvius mons Campaniae torridus intestinis ignibus aestuans exusta evomuit viscera nocturnisque in die tenebris incumbentibus omnem Europae faciem minuto contexit pulvere. Huius metuendi memoriam cineris Byzantii annue celebrant VIII idus Novemb* (»[…] the scorched mountain of Campania, seething with internal fires, spat out its scorched contents that obscured the daylight, and covered the whole surface of Europe with fine dust. This terrible dust is commemorated every year on November 6 in Byzantium.«), transl. Croke, 25. Cf. Cassiodorus, *Variae* IV, epist. 50, 4, ed. Mommsen 137: *Volat per inane magnum cinis de coctus et terrenis nubibus excitatis transmarinas quoque provincias pulvereis guttis compluit, et quid Campania pati possit, agnoscitur, quando malum eius in orbis alia parte sentitur.* (»Then the air is darkened by its foul vapors, hot ash rushes along the sea, a shower of dust drops [comes] over the country and reports to all of Italy and to the overseas provinces in the world of the misfortune Campania suffers.«). Rolandi et al., The A.D. 472 eruption of the Somma volcano, 293-294, 301-307; Wozniak, *Naturereignisse im frühen Mittelalter*, 325-326.

144  Ioannes Malalas, *Chronographia* XIV 42, ed. Thurn, 295; *Chronicon Paschale*, ed. Dindorf, vol. 1, 598, 10-14; Grey, Risk and vulnerability, 3, 25-26, 33. On both sources, see Hunger, *Die hochsprachliche profane Literatur der Byzantiner*, 319-326 and 328-330; Neville, *Guide to Byzantine Historical Writing*, 52-55 (no reference to Malalas).

people wailed and lamented piteously, fearing that a flood like that fabled one of old was again befalling them. But compassionate Providence, which loves mankind, thrust a rainbow through the clouds, and with its rays dispersed the gloomy rain, and the structure of nature returned again to its previous condition. It so happened that there was a later downpour, which was turbid and mixed with ashes (in Greek, tephra), as in the soot from a furnace, and it seemed lukewarm to those who touched it.[145]

Furthermore, on 22 December 968, »an eclipse of the sun took place«, so that once more »people were terrified at the novel and unaccustomed sight, and propitiated the divinity with supplications, as was fitting«. As Leo does not forget to mention, he was an eyewitness, since »at that time I myself was living in Byzantium, pursuing my general education.«[146]

The »downpour (…) mixed with ashes« could have been the result of an eruption of Mt. Vesuvius in 968; according to the Chronicle of Montecassino, in this year, »the Vesuvius exploded with flames and ejected a great mass of viscous and sulfurous substance, which formed a stream flowing rapidly into the sea.«[147] The significant magnitude of this eruption of 968 was also reconstructed on the basis of geological evidence.[148]

The solar eclipse of December 968 is equally mentioned by Liutprand of Cremona towards the end of his description of his second embassy (»relatio de legatione«) to Constantinople.[149] Wozniak, however, claims with reference to the above-mentioned NASA database that no eclipse took place around that time. He therefore assumes that the darkening of the sun described by Leo the Deacon and Liutprand was also caused by the ashes of Mt. Vesuvius.[150] A check in the NASA database, though, yields a clear reference to a solar eclipse on 22 December 968, whose umbra shadow exactly covered Constantinople and regions to the southwest and northeast of the capital.[151] The solar phenomenon mentioned by Leo the Deacon and Liutprand thus does not need to be connected with the Vesuvius eruption of 968, which, however, may have caused the downpour of ashes in Constantinople in June of the same year.

The ashfall of 472 CE is also registered in the *Chronographia* (under the year AM 5966)[152], but without any reference to Mt. Vesuvius as source of this phenomenon:

In this year dust came down from clouds that seemed to be burning, so that everyone thought it was raining fire. Everybody performed litanies in fear. The dust settled on roofs to the depth of one palm. Everybody said that it was fire and that it was put out and became dust through God´s mercy.[153]

---

By contrast, a large volcanic eruption in (spatial and temporal) proximity to Theophanes is described in relative detail for the year 726 CE, the one of the volcano of Thera (Santorini) in the Southern Aegean (see the map Fig. 16), whose effects (in the form of downfalls of pumice stone) could be felt as far as Asia Minor, Lesbos, Macedonia and the Dardanelles (Abydos is mentioned) – and according to the *Chronographia* even motivated Emperor Leon III to initiate »a more ruthless war on the holy and venerable icons«.[154] A parallel description (and connection with the »iconoclast« policy of Leon III) is provided by Patriarch Nikephoros.[155]

For the darkening of the sun in August 797, we find no reference of this kind in the *Chronographia* which could provide additional indications for a volcanic eruption in the Mediterranean environs of Constantinople. We get information on volcanic activities of Mt. Vesuvius and other regions of Italy during the reign of Eirene and Constantine VI, however, from another source, namely an eyewitness report by a certain Gregorios,[156] registered in the *Synaxarion* of Constantinople for 30 April in connection with notices on saints Patrikios and Pionios:

If anyone does not believe it [the stories just told on the saints Patrikios and Pionios], consider the island of Lipari which is subject to fire so much so that it makes the sea boil, to swallow the ships that are there, while the spruce lava flows liquefied, and tremendous thunder is produced from that islet. And then all of Lipari is shaken and trembles; the dune of the sea rises all on fire from the depths and rises to infinite heights, and is carried by any wind by fate, and falls here and there. Some still say this, that when it is known that some wicked and iniquitous passed away from life, then those places suffer eruptions of fire and thunder, as if those souls are condemned to punishment there. For at these places also I, Gregory, traveling around after the second sacred Synod in Nicaea had happened, heard and saw those wonderful things. And again, when I arrived in Naples while we were traveling by sea towards the older Rome, I saw in Naples itself that mountain that is six miles from the city and over- looks it [the Vesuvius], and it is all cavernous, as if it were throwing divine fire, as if the waters were gushing from its top. And that fire went down to six miles, so that by flooding for six days it burned the earth and the stones and the stone buildings and the plants, and reduced everything to ashes, until Stephanos,[157] who was then the most holy bishop there (Στέφανος ὁ τηνικαῦτα ὁσιώτατος ἐπίσκοπος), came out with a de-voted procession of supplicants, came near the fire and prayed, and the wrath of God was appeased and stopped. However, when the eruption was alive, huge boulders rose from the ground in the middle of the fire, and were thrown to immeasurable heights.  In the day there was a great column of smoke raised up to the sky, and in the night that column was of fire. Such things God makes men see in order to reduce them to change their minds, so that by abandoning the ways of iniquity, and placing themselves on the paths of health they may come to possession of the kingdom of heaven, which we all, as we hope, will come to possess, and so be it.[158]

---

A terminus post quem for the eruption of Mt. Vesuvius observed by Gregorios is provided by the reference to the Second Council of Nicaea, which took place between 24 Septem- ber and 23 October 787; therefore, the year 787 is presented as the date for this eruption in several studies.[159] Michael McCormick, however, assumed that Gregorios traveled back from Nicaea (most probably via Constantinople) only in spring 788. Also, in two catalogs of volcanic eruptions, the event is therefore dated between October 787 and March 788.[160] The wording of Gregorios further indicates that Stephanos, the bishop of Naples, had died by the time he wrote his text. The death of Stephanos in 800 provides a firm terminus ante quem for Gregorios' journey. If we assume that he may have stayed even longer in Byzantium after the council in Nicaea, the widest possible dating interval for the eruption of Mt. Vesuvius he observed is between autumn 787 and the year 800, although the direct reference to the council suggests a dating nearer to the first year than to the latter.[161]

As usual, natural scientific dating provides an even wider temporal framework; calibrated $C^{14}$ dates of samples of charcoal and paleosols which could be connected to the eruption of Mt. Vesuvius described by Gregorios range between AD 690/775/780 and 935/980/1017.[162] The archaeomagnetic age of lava at the site of Masseria Galassi, along the coast, near Torre Annunziata is indicated with AD 800 +/- 20 years.[163] Although other contemporary writ- ten evidence is missing, based on Gregorios' description and the geological evidence, the

»787 eruption« has been described as a significant event »of the mixed (effusive and ex- plosive) type«, marking the beginning of a new period of increased volcanic activity on Mt. Vesuvius.[164] Reconstructions of the diffusion of volcanic ash from Mt. Vesuvius as well as computer models based on modern-day air circulation data show that there was and is a very high probability that aerosols and gases ejected from Mt. Vesuvius after eruptions on such a scale found and find their way to the east towards the Aegean and the Bosporus.[165]

Previous to Mt. Vesuvius, however, Gregorios mentions volcanic activity on the Aeolian or Liparic islands, which he passed by on a ship on his journey back from Byzantium. We also hear about eruptions there in the late 7th and early 8th centuries in two pilgrimage reports on the Holy Land, by the Irishman Adomnanus (receiving most of his information from the Frankish bishop Arculf, who traveled to the East around 680) and the Anglo-Saxon mission- ary Willibald (who traveled to Jerusalem before 727)[166] — by each in the description of the

---

return journey to the West.[167] Two motivations may have been decisive for reporting such natural events, which do not fit within the places of Christian salvation that were otherwise visited and described. On the one hand, the main shipping route at that time led from the east through the Strait of Messina and then passed the Aeolian Islands on the way north.[168] On the other hand, the Aeolian island of Vulcano had the reputation that its main crater was an entrance to the underworld, already in relation to the ancient pagan god Hephaestus/Vulcanus as lord of the underworld. The Christian legend that the Ostrogothic king Theodoric, who was considered a heretic because of his Arian confession, had gone to hell through the crater on Vulcano after his death in 526, had a reinforcing effect on this very idea.[169] Eruptive ejec- tion of pumice (primarily from Monte Pilato on Lipari), the shape of which sometimes re- sembled human body parts,[170] was interpreted to mean that the guilt-ridden deceased atoned for their sins in volcanic (hellish) fire (to which the blazing flames of the craters testified, clearly visible at night) – see also the description by Gregorios cited above.[171]

Coming back to the narrative by Gregorios, chronologically close and potentially connected with the volcanic activity described by him for Lipari in the late 780s or 790s (see the map Fig. 16) is the so-called »Monte Pilato Event«, whose geological remains have a calibrated C[14] date of AD 776 + 110/-90 years (respectively AD 780-785 in an earlier study).[172] The date range of the C[14] method would thus allow us to identify this eruption of Monte Pilato on Lipari with the one described by Gregorios for the years between 787 and 800 – at least, Gregorios´ observations provide a parallel confirmation for a period of increased volcanic ac- tivity on Lipari in the late 8th century.[173] The date range for the Monte Pilato Event is further confirmed by the dating of geological traces of the eruption on neighboring Aeolian islands such as Vulcano, where near-contemporaneous eruptions also occurred.[174]

---

Based on the geological evidence, the Monte Pilato Event is described as »a rather strong explosive activity«[175] of »at least sub-Plinian character«[176] which also had effects on human settlement on the island of Lipari.[177] Ashes from the Monte Pilato eruption(s) of the late 8th century have been found in the Gulf of Taranto, 250 km to the east.[178] The wider diffusion patterns of volcanic material from the Aeolian Islands are similar to the one described above for Mt. Vesuvius, with a strong axis towards the east, i.e. the Ionian Sea and the southern Balkan Peninsula.[179]

Both written evidence and geological data document significant volcanic activity on Mt. Vesuvius and Lipari in the years between 787 and 800; earlier and later historical sources as well as modern-day reconstructions show that eruptions on such as scale could cause atmos- pheric phenomena all the way to Constantinople (while not affecting Central and Western Europe, which could explain the above-discussed lack of parallel reports to the one in the *Chronographia* for August 797).

As mentioned above, the possible dating interval for the eruptions of Mt. Vesuvius and  on Lipari observed by Gregorios is between autumn 787 and the year 800; the linkage to Gregorios´ journey to the Second Council of Nicaea, however, suggests a dating nearer to 787 than to 800. So how could the volcanic activities described by Gregorios relate to the atmospheric phenomena mentioned by the *Chronographia* for August 797?

As mentioned several times, the *Chronographia* was quite »flexible« in the distribution of its material across the years of the chronicle, especially if this supported narrative strategies. This is also true for the entry on AM 6289 (1 September 796-31 August 797), which mentions the blinding of Emperor Constantine VI (see above), or the above-mentioned solar eclipse in 787, which the *Chronographia* registered for AM 6279 (1 September 786-31 August 787) in order to reserve AM 6280 (1 September 787-31 August 788) for the Second Council of Nicaea and to keep this year free from any portents which could have derogated this event, such as a solar eclipse.

Accordingly, the darkening of the sun described for August 797 as a »celestial« reaction to the blinding of Constantine VI could have taken place earlier (or later). If we assume an early date of the eruptions described by Gregorios in either late 787 or early 788, as for the eclipse of September 787, the *Chronographia* could have decided to exclude a resulting atmospheric phenomenon in Constantinople from the correct AM 6280, reserved for the Second Coun- cil of Nicaea, and instead to use the portent for framing the blinding of the emperor in 797. Alternatively, Gregorios´ journey past Lipari and Mt. Vesuvius and the described vol- canic eruptions could have taken place some years later in closer temporal proximity to 797, which would have made it even easier (or tempting) for the *Chronographia* to re-arrange the chronology of the events. Thus, the volcanic activity of Mt. Vesuvius and on Lipari provides another potential geophysical background to the atmospheric phenomena described by the *Chronographia* in conjunction with the blinding of Emperor Constantine VI.

---

*Conclusion*

In summary, we can state that the combination of an informed reading of the written sources with the findings of the natural sciences enables an evaluation of the assumptions made so far about the background of the 17-day darkening of the sun mentioned by the *Chrono-graphia* for August 797 in connection with the blinding of Emperor Constantine VI. The assumption that the chroniclers linked a solar eclipse to the mutilation of the emperor is to be rejected. A highly probable cause of the phenomena described (darkening of the sun for more than two weeks, obstruction of navigation due to poor visibility) is a turbidity of the atmosphere caused by volcanic aerosols, as described by eyewitnesses in the wake of other eruptions in a manner comparable to the *Chronographia*.

However, the exact date and the location of this volcanic event remain unclear. Various hypotheses about a mention of atmospheric turbidity in Frankish sources at the same time as 797/798 appear to be insupportable, as do earlier speculations that suggest Mount Churchill in Alaska as the volcano responsible (at least in connection with the eruption responsible for the dispersal of the White River Ash).

The evidence for the incidence of a notable volcanic eruption in the year 799/800, documented by (revised) ice core data from Greenland and by data (the tree rings in Scandinavia and the Swiss Alps and perhaps also in the reports on weather anomalies in 800/801 in Frankish sources) from other parts of Europe on its likely climatic effects, in contrast, is strong. The same is true for eruptions of Mt. Vesuvius and on Lipari (both documented in written and geological evidence) at some time after the Second Council of Nicaea, which took place in autumn 787.

Thus, based upon our review of the available evidence, we may hypothesize that the *Chrono- graphia* (maybe following »public memory« in Constantinople) either combined an earlier veiling of the atmosphere apparent over the Bosporus due to the eruption of Mt. Vesuvius or an eruption on Lipari with its account of the blinding of Constantine VI in August 797, or with the veiling caused by a later major eruption around the year 799/800 (an overview of the events and phenomena discussed in our paper can be found in Table 2). For all these volcanic events, the range of geological age determination at least would even allow arguing for a dating to the late summer of 797, but the written evidence for Mt. Vesuvius and Lipari suggests a date closer to 787/788. For the later major eruption, tree rings (and the references in the Frankish historiography) suggest a date of c. 799/800. It remains to be seen if further refinements of the dating of volcanic events based upon new natural or written evidence for the last years of the 8th century will allow for a more direct source attribution to be made to the August 797 darkness without resorting to the (however highly probable) scenario of an adaptation of the actual date of this darkening to suit the narrative needs the *Chronographia*.

In finally returning to consider the historiography of these important years, we can certainly state that the atmospheric phenomena and climatic disturbances observed (and reconstructed from recent data) around the year 800 allowed for a peculiar framing of the political events in the eyes of contemporary authors, who within a few years were confronted with the spectacular fall and rise of emperors at the Bosporus and the Tiber.[180] As Jesse W. Torgerson states, the »*Chronographia* was written to tell the truths of the past for its own present, not ours«,[181] and this truth also included a portentous reading of celestial, atmospheric and cli- matic phenomena which may seem unfamiliar to the »enlightened« modern-day reader.

---

180  Most recently on this context, see Kislinger, Diskretion bis Verschleierung.

181  Torgerson, *Chronographia*, 395.

However, a critical reading and combination of written sources and the growing number of natural archives allows us to connect these interpretations with specific natural dynamics – and in turn to illustrate that as historians, we have to reckon not only with the sheer phys- ical impact (such as damages) of these phenomena on socio-political developments, but also with their interplay with the interpretation of such developments within respective cultural frameworks.

| Time | Event or phenomenon | (Possible) physical background |
|---|---|---|
| 726 | Major volcanic eruption on Thera (Santorini) in the Southern Aegean | |
| 746, March-April | Veiling of the sun in Syria and Meso- potamia | Atmospheric turbidity (after a volcanic eruption or dust storms) |
| 747 or 748, April | Birth of Charlemagne | |
| 747, August | Veiling of the sun in Syria and Meso- potamia for 5 days | Atmospheric turbidity (after a volcanic eruption or dust storms) |
| 747/747 | Last outbreaks of the first plague pandemic in Constantinople and in the Caliphate | |
| ca. 752 | Birth of Eirene | |
| 762/764 | Sulfate peak in ice cores in Greenland | Major volcanic eruption |
| 763-764, winter | Extreme cold winter across Europe, freezing of parts of the Black Sea and the Bosporus around Constantinople | Climate anomaly after a volcanic eruption |
| 764, March | »Falling of stars« in Constantinople | Meteorite shower? |
| 767, Summer | Severe drought around Constantinople | |
| 768, 9 October | Accession to power of Charlemagne in the Frankish Kingdom | |
| 769 | Wedding of Eirene to Leon (IV), son of Emperor Constantine V | |
| 771, 14 January | Birth of Constantine VI | |
| 773/774 | Charlemagne conquers the Lombard Kingdom in Northern Italy | |
| 774/775 | Sightings of auroras a far south as Amida in northern Mesopotamia | Massive outbreak of the sun (»774/775 event«) |
| 775, 14 September | Death of Emperor Constantine V, accession to the throne of Leon IV | |
| 780, 8 September | Death of Emperor Leon IV, acces- sion to the throne of Constantine VI and of Eirene (as co-empress) | |

| Time | Event or phenomenon | (Possible) physical background |
|------|---------------------|-------------------------------|
| 787, 16 September | Solar eclipse, partially visible in Constantinople | |
| 787, 24 September-23 October | Second Council of Nicaea, presided over by Constantine VI and Eirene | |
| 787/800 | Volcanic eruptions of Vesuvius near Naples and Monte Pilato on the Liparic Islands | |
| 790, February | Discord of Constantine VI and Eirene, earthquake in Constantinople | |
| 790, October | Constantine VI has Eirene confined to the Palace of Eleutherios, fire in Constantinople | |
| 792, August | Emperor Constantine VI orders the blinding of his paternal uncles after an attempted coup | |
| 792, 25 December | Rebellion of the military corps of the Armeniacs against Constantine VI, fire in Constantinople | |
| 792-794 | Periods of bad weather and crop failures in the Frankish realms | Climate anomaly (maybe after a volcanic eruption?) |
| 795, September | Constantine VI separates from his first wife Maria and marries Theodote | |
| 796, April-May | Earthquakes in Crete and later in Constantinople | |
| ca. 796 | Ammonium peak in ice cores in Greenland | Large-scale wildfires in North America? |
| 797, 3 March | Solar eclipse, partially visible in Constantinople | |
| 797, July-798, July | Planet Mars not visible on the night sky in the Frankish realms | Conjunction of Mars |
| 797, August | Blinding of Emperor Constantine VI in Constantinople, »darkening of the sun« for 17 days | Atmospheric turbidity after a volcanic eruption |
| 799, 25 April | Blinding of Pope Leo III in Rome | |
| 799/800 | Sulfate peak in ice cores in Greenland | Major volcanic eruption |
| 800 | Severe cold summer registered in tree rings in Scandinavia and Switzerland, frost in June in the Rhineland, extreme drought in the Empire of the Uyghurs (Mongolia) | Climate anomaly after a volcanic eruption |

| Time | Event or phenomenon | (Possible) physical background |
|---|---|---|
| 800, 25 December | Coronation of Charlemagne as Emperor of the Romans in Rome by Pope Leo III | |
| 800-801, winter | Storms in England, mild winter in the Rhineland, severe winter in Catalonia | Climate anomaly after a volcanic eruption |
| 801, spring | Epidemics among cattle and humans in England and the Rhineland | |
| 802, October | Overthrow of Empress Eirene by Nikephoros I, bad weather | |
| 803, 9 August | Death of Eirene in exile on Lesbos | |
| 809/810 | Severe epidemic among cattle and horses in the Carolingian Empire | |
| 812, 14 May | Solar eclipse, partially visible in Constantinople | |
| 813, 4 May | Solar eclipse, partially visible in Constantinople | |
| 814, 28 January | Death of Charlemagne in Aachen | |
| 822 | Sulfate peak in ice cores in Greenland, extreme weather in Ireland | Climate anomaly after a volcanic eruption (Katla on Iceland) |

*Table 2: Selected political events, portents and natural phenomena in the 8th to 9th century AD*

*Acknowledgements*
We would like to thank the two anonymous reviewers for their most valuable comments and additions, as well as Michael Sigl (Physics Institute, University of Bern) and Hisashi Hayakawa (Institute for Space-Earth Environmental Research, Nagoya University) for their indispensable advice.

*Websites*

*List of Figures*

Fig. 7: Umbra shadow (blue) and zone of partial visibility (green) of the to-
tal solar eclipse of 4 May 813 (Image source accessed on 21 May 2021:
eclipse.gsfc.nasa.gov/5MCSEmap/0801-0900/813-05-04.gif)

Fig. 8: Expansion of the ash cloud after the eruptive eruption of Eyjafjallajökull in Ice-
land on 14 April 2010, until 18 April 2010 (Image source accessed on 21 May 2021:
commons.wikimedia.org/wiki/File:Eyjafjallaj%C3%B6kull_volcanic_ash_18_April_2010.png)

Fig. 9: The expansion of the »Höhenrauch« (haze) after the eruption of the Laki volcano in
Iceland on 8 June 1783, during the following summer according to eyewitnesses (adapted
from: Mikhail, Ottoman Iceland; map: J. Preiser-Kapeller, 2022)

Fig. 10: The southern starry sky in the summer months in Central Europe with the constella-
tion Scorpius (in German »Skorpion«) and the star Antares just above the horizon (source
accessed on 21 May 2021: astrokramkiste.de/himmel-im-sommer)

Fig. 11: Tree ring data from the Lötschental in Switzerland for the period 750 to 850 (data:
Guillet *et al.*, Climatic and societal impacts; graphic: J. Preiser-Kapeller, 2022)

Fig. 12: **(Top)**: NEEM(S1) Greenland ice core: non-sea-salt-sulfur concentrations on the NS1-
2011 chronology (constrained with 775 CE 10Be/14C anomaly).[182] Red circles are
eruptions attributed on the basis of geochemistry of tephra shards extracted from ice
core records in Greenland. **(Middle)**: the time window 775-860 CE from the NEEM(S1)
Greenland ice core data shown at the top. (**Bottom**): the time window 775-860 CE from
the TUNU2013 Greenland ice core record. The Katla 822/823 eruption is dated by
dendrochronology[183] and tephra was identified in the TUNU2013 ice core.[184] The Mt.
Churchill eruption (White River Ash B) was identified by tephra in the NEEM and NGRIP
ice cores[185] (Graphs by courtesy of Michael Sigl).

Fig. 13: Tree ring data from Torneträsk in Sweden for the period 750 to 850 (data: Guillet
*et al.*, Climatic and societal impacts; graphic: J. Preiser-Kapeller, 2022)

Fig. 14: Tree ring data from northern Sweden and Finland for the period 750 to 850 (data:
Guillet *et al.*, Climatic and societal impacts; graphic: J. Preiser-Kapeller, 2022)

Fig. 15: Reconstruction of summer temperatures in the Northern Hemisphere for the period
500 to 2003 (compared to the period 1961-1990) based on tree ring data from
different regions (data: Guillet *et al.*, Climatic and societal impacts; graphic: J. Preiser-
Kapeller, 2022)

Fig. 16: Map of selected historical and natural scientific evidence discussed in the paper
for the late 8th and early 9th centuries CE (map: J. Preiser-Kapeller, 2022)

---